\documentclass{ar2e}

\usepackage{graphicx}
\usepackage{amsmath}
\usepackage{bm}
\usepackage{cite}

\newcommand{\bqa}{\begin{eqnarray}}
\newcommand{\eqa}{\end{eqnarray}}
\newcommand{\be}{\begin{equation}}
\newcommand{\ee}{\end{equation}}
\newcommand{\ba}{\begin{eqnarray}}
\newcommand{\ea}{\end{eqnarray}}
\newcommand{\beq}{\begin{equation}}
\newcommand{\eeq}{\end{equation}}
\newcommand{\beqa}{\begin{eqnarray}}
\newcommand{\eeqa}{\end{eqnarray}}
\newcommand{\fet}[1]{\mbox{\boldmath $#1$}}
\newcommand{\nn}{\nonumber \\ }

\begin{document}



\jname{Annual Review in Nuclear and Particle Science}
\jyear{}
\jvol{}
\ARinfo{}

\title{\baselineskip 18pt
Chiral dynamics of few- and\\many-nucleon 
systems}

\markboth{Epelbaum, Mei{\ss}ner}{Chiral dynamics of few- and many-nucleon 
systems}

\author{Evgeny Epelbaum\footnote{Corresponding author}
\affiliation{\baselineskip 13 pt
Institut f\"ur Theoretische Physik II, Ruhr-Universit\"at Bochum,\\
  D-44780 Bochum, Germany\\
        Email: evgeny.epelbaum@rub.de}
Ulf-G.~Mei{\ss}ner
\affiliation{\baselineskip 13 pt
Universit\"at Bonn, Helmholtz-Institut f{\"u}r
  Strahlen- und Kernphysik and\\ Bethe Center for Theoretical Physics, 
  D-53115 Bonn, Germany; \\
Forschungszentrum J\"ulich, Institut f\"ur Kernphysik 
  (IKP-3), Institute for Advanced Simulation (IAS-4) and J\"ulich Center
   for Hadron Physics,\\  D-52425 J\"ulich, Germany\\
Email: meissner@hiskp.uni-bonn.de}
}

\begin{keywords}
\baselineskip 13 pt
QCD, chiral symmetry, effective field theory,  nuclear
forces, lattice simulations, nuclear physics 
\end{keywords}

\begin{abstract}
\baselineskip 14pt 

This review gives a brief introduction to the chiral effective field 
theory of nuclear forces and atomic nuclei. We discuss the status of the
nuclear Hamiltonian derived in this framework and some recent applications
in few-nucleon systems. Nuclear lattice simulations as a new tool to address
the many-body problem are introduced and some first results based on
that method are presented.

\noindent\footnotesize{
\begin{center}Commissioned article for {\it Ann. Rev. Nucl. Part. Sci}
\newpage
\end{center}}
\end{abstract}

\maketitle
\topmargin 2cm

\baselineskip 18pt 
\newpage
\section{Introduction and disclaimer}

Nuclear physics is  on one side an old and well-established science
but on the other hand a fascinating and new field. This is related to new
experimental facilities and techniques but even more so to recent developments
in theory. Using modern high-performance computers, first attempts are
made to calculate atomic nuclei directly from Quantum Chromodynamics (QCD),
the SU(3)$_{\rm color}$ gauge theory of quarks and gluons
\cite{Beane:2010em}. 
Complementary to this, starting from the ground-breaking work of Weinberg
\cite{UTTG-31-90}, an effective field theory (EFT) 
approach to the forces
between two, three and four nucleons has been developed and applied to a
variety of nuclear bound states and reactions. This EFT is based on the
observations that (i) nuclei are composed of non-relativistic nucleons 
(neutrons and protons) and
virtual mesons and (ii) that the nuclear interactions feature two very distinct
contributions, long-range one- and two-pion exchanges and shorter-range
interactions, that can be represented by a tower of multi-nucleon operators.
As the pion is the pseudo-Goldstone boson of the approximate chiral symmetry
of QCD (for an introduction, see e.g. Ref.~\cite{Bernard:2006gx}), its interactions with  the 
nucleons are of derivative nature and strongly constrained by the available
data on pion-nucleon scattering and other fundamental processes. However,
in harmony with the principles underlying EFT  (for an introduction, see
e.g. Ref.~\cite{Georgi:1994qn}), 
one has also to consider operators of nucleon fields only. 
In a meson-exchange model of the nuclear forces, these can be pictured 
by the exchanges of heavier mesons like $\sigma$, $\rho$, $\omega$, and 
so on - but such a modeling is no longer necessary and also does not 
automatically generate all structures consistent with the underlying
symmetries. Also, in the EFT approach, the forces between three and four
nucleons are generated consistently with the dominant two-nucleon forces -
which could never be achieved in earlier modeling of these forces.

Due to the non-relativistic nature
of nuclei, the underlying equation for the nuclear $A$-body system (where
$A$ is the atomic number) to be solved is the Schr\"odinger 
equation, where the various contributions to the nuclear
potential are organized according to the power counting discussed below. In a second
step, bound and scattering states are calculated as solutions of this
equation. This does not only allow to pin down the various low-energy
constants related to the multi-nucleon interactions, but also to check the
convergence of the approach by including higher orders in the underlying
potentials. Once data for the two- and three-nucleon systems are described
with sufficient precision, one is then in the position to perform {\sl ab 
initio} calculations of nuclei, eventually combining well developed many-body
techniques with the forces from chiral nuclear  EFT. Another venue to approach
light and medium-heavy nuclei is based on simulation techniques, as they are
so successfully utilized in lattice QCD \cite{Montvay:1994cy} to calculate the properties of
protons, neutrons and many other hadrons. All this is accompanied and extended
by the construction of the corresponding electro-weak charge and current 
operators that allow for many further fine tests of the structure of nuclei
and also of the calculation of fundamental nuclear reactions that are of relevance
to the generation of the elements in the Big Bang and in stars.

{\sl Disclaimer:}~Clearly, this is {\sl not} a detailed all purpose review of this field but
it rather intends to give an introduction to the underlying ideas and some recent
applications. Two recent detailed review articles are 
Refs.~\cite{Epelbaum:2008ga,Machleidt:2011zz}
that contain also many references to earlier work. We have therefore not
attempted to be complete or exhaustive in the references, but the interested
reader will find sufficient quotes to the literature to be able to
acquire a much deeper understanding. 

Our article is organized as follows: Section~2 contains the basic ideas of
the chiral Lagrangian and power counting of the nuclear forces. We end this
section with a brief review of the current status of the nuclear Hamiltonian
derived in this framework. Section~3 contains some applications of these forces 
to nuclei, based on calculations using exact few-nucleon
methods.
In section~4, the new method of nuclear
lattice simulations (nuclear lattice EFT) is presented and some first results
obtained in that scheme are displayed.

\section{From the effective chiral Lagrangian to nuclear forces} 

\begin{figure}[t!]
\centerline{\includegraphics[width=0.6\textwidth,keepaspectratio,angle=0,clip]{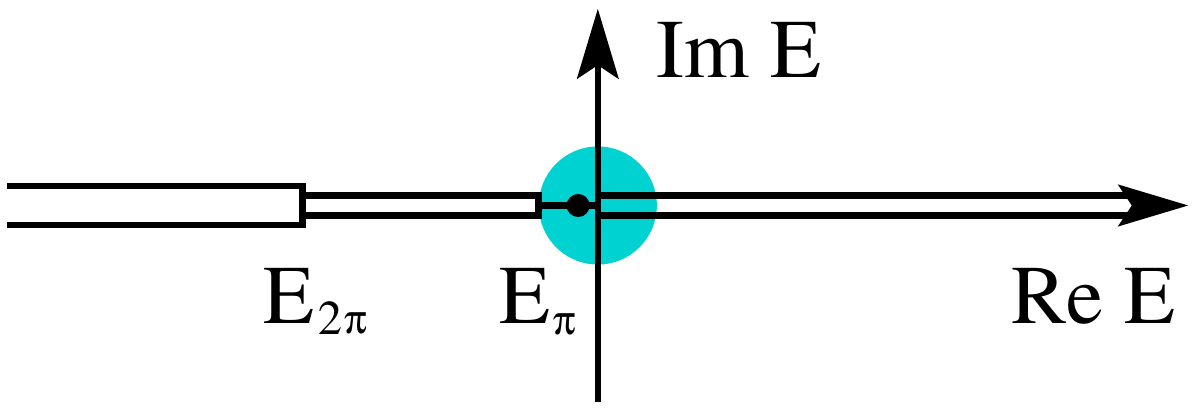}}
    \caption{\baselineskip 12pt 
         Singularity structure of the partial-wave two-nucleon
         scattering amplitude in the complex energy plane. The solid
         dot indicates the position of the S-wave (virtual) bound
         state. The shaded area shows the region where the effective
         range function $k^{2l+1} \cot \delta_l (k)$ is a meromorphic function of $k^2$. 
\label{fig1} 
 }
\end{figure}

Our goal is to develop a systematic and model-independent theoretical framework capable to describe 
reactions involving several nucleons up to center-of-mass three-momenta of (at
least) the 
order of the pion mass $M_\pi$. Following the usual philosophy of effective field theory, we aim 
at the most general parameterization of the amplitude consistent with the fundamental principles 
such as Lorentz invariance, cluster separability and analyticity. Given that the energies of
the nucleons we are interested in are well below the nucleon mass, it is
natural and 
appropriate to make use of 
the non-relativistic expansion (i.e.~an expansion in inverse powers of the nucleon mass $m_N$). 
Accordingly, in the absence of external probes and below the pion production 
threshold, we are left with a potential theory in the framework of 
the quantum-mechanical $A$-body Schr\"odinger equation
\beq
\label{SE}
\big( H_0  + V \big) |
\Psi \rangle = E | \Psi \rangle\,,  \quad 
\quad H_0 = \sum_{i=1}^A \frac{\vec \nabla_i^2}{2
  m_N}  + \mathcal O (m_N^{-3})\,.
\eeq
The main task then reduces to the determination of the nuclear
Hamilton operator $H_0 +  V$. This can be 
accomplished using the framework of chiral perturbation theory (ChPT) \cite{UTTG-31-90}.  
Notice that the approach outlined above automatically maintains unitarity of the scattering amplitude 
and correctly reproduces its analytic properties at very low energies. 
Consider, for example, the singularities 
of the neutron-proton $l^{\rm th}$ partial wave amplitude in the complex
energy plane with $E=k^2/m_N$, visualized schematically in Fig.~\ref{fig1}.
The discontinuity across the 
right-hand cut running from $E=0$ to $E= + \infty$ and separating the physical and unphysical 
sheets is determined by elastic unitarity which is already built in
the Lippmann-Schwinger equation. On the other hand,
the left-hand cuts are governed by the properties of the interactions.  
Exploiting only the knowledge of the finite-range nature of the nuclear force and parameterizing the nucleon-nucleon (NN)
potential by zero-range terms is sufficient to correctly describe the analytic structure of the amplitude
within the region near threshold limited by the branch point on the
first left-hand cut associated with one-pion exchange and 
located at $k^2 = M_\pi^2/4$. 
The so-called pionless EFT, see \cite{arXiv:0910.0031} for a recent
overview, is based on the Lagrangian involving all possible zero-range
two- and more-nucleon  
operators with increasing number of derivatives and can, in the two-nucleon sector, be matched to 
the well-known effective-range expansion, i.e.~an expansion of the function $k^{2l+1} \cot \delta_l (k)$ in 
powers of $k^2$. Clearly, this approach is
limited to very low energies corresponding to nucleon momenta  
well below the pion mass, see Fig.~\ref{fig1}. To extend the range of
applicability,  one needs to correctly describe the left-hand 
singularities of the amplitude and thus to explicitly include in the potential the contributions emerging from 
the exchange of one or several pions.  We will describe in the next subsections how this task can be accomplished 
in a systematic way by exploiting the spontaneously broken, approximate chiral
symmetry of QCD. Another method to construct the NN scattering amplitude is
dispersion theory, see 
Ref.~\cite{Albaladejo:2011bu} for a recent application of this approach.

\subsection{Chiral Lagrangian and power counting} 

Within the framework of ChPT, nuclear forces are 
derived from the most general effective chiral Lagrangian by making
an expansion in powers of the small parameter $q$ defined
as\footnote{\baselineskip 12pt Note that we use this parameter and the
soft scale $Q$ synonymously in what follows.}
\beq
\label{deltaless}
q \in \bigg\{ \frac{M_\pi}{\Lambda}, \,    \frac{| \vec k\, | }{\Lambda} \bigg\}\,,
\eeq
where $Q \sim | \vec k | \sim M_\pi$ is a typical external momentum (the soft
scale) and $\Lambda$ 
is a hard scale. Appropriate powers of the inverse of this scale determine the
size of 
the renormalized low-energy constants (LECs) in the effective Lagrangian.  
Notice that once renormalization of loop contributions is carried out and the renormalization scale 
is set to be $\mu \sim M_\pi$ as appropriate in ChPT, all momenta flowing
through diagrams appear to be, effectively, of the order $ \sim M_\pi$
\cite{HUTP-78-A051A}. Consequently, one can use naive dimensional analysis 
to estimate the importance of (renormalized) contributions of
individual diagrams. 

To be specific, consider a connected 
Feynman graph with $N$ nucleon lines.\footnote{\baselineskip 12pt
We remind the reader that nucleons cannot be destroyed 
or created within the nonrelativistic approach.} It is easier to count the powers 
of the hard scale $\Lambda$ rather then of the soft scale $Q$ by observing 
that the only way for $\Lambda$
to emerge is through the corresponding LECs. 
Thus, the 
low-momentum dimension $\nu$ of a given diagram can be expressed in terms of the canonical 
field dimensions $\kappa_i+4$  of $V_i$ vertices of type $i$  via
\beq
\label{pow_co_alg}
\nu = -2 + \sum V_i \kappa_i \,, \quad \quad \kappa_i = d_i + \frac{3}{2} n_i + p_i -4 \,,
\eeq
where $n_i$ ($p_i$) and $d_i$ refer to the number of the nucleon (pion) field operators and derivatives or 
pion mass insertions, respectively. The constant $-2$ in the expression for
$\nu$ is just a convention.   
The power counting can also be re-written in terms of topological variables such as the number of 
loops $L$ and nucleon lines $N$ rather than $\kappa_i$ which are appropriate for diagrammatic approaches. 
 For connected diagrams the above equation then takes the form  
\beq
\label{powc_w}
\nu = -4 + 2N + 2 L + \sum V_i \Delta_i \,, \quad \quad \Delta_i = d_i + \frac{1}{2} n_i -2 \,.
\eeq
Chiral symmetry
of QCD guarantees that Goldstone Bosons which, in the case of two
light flavors are identified with the pions, couple only through vertices
involving derivatives or powers of $M_\pi$.  This implies 
that the effective Lagrangian contains only irrelevant (i.e.~non-renormalizable) interactions with 
$\kappa_i \geq 1$ ($\Delta_i \geq 0$) which allows for a
\emph{perturbative} description of pion-pion and pion-nucleon scattering
as well as nuclear forces. 
The leading interactions, i.e.~the ones  with the smallest possible
$\Delta_i$, that is $\Delta_i=0$,  have the form 
\beqa
\label{lagr}
 \mathcal{L}^{(0)} &=& \frac{1}{2} \partial_\mu \fet \pi \cdot \partial^\mu \fet \pi - 
\frac{1}{2}  M_\pi^2 \fet \pi^2 + N^\dagger \left[ i \partial_0 + \frac{g_A}{2 F_\pi} \fet \tau \vec \sigma 
\cdot \vec \nabla \fet \pi - \frac{1}{4 F_\pi^2} \fet \tau \cdot (\fet \pi \times \dot{\fet \pi}) \right] N\nn
&& {} - \frac{1}{2} C_S (N^\dagger N) (N^\dagger N) -  \frac{1}{2} C_T (N^\dagger \vec \sigma N) \cdot (N^\dagger 
\vec \sigma N) + \ldots \,, 
\eeqa
where $\fet \pi$ and $N$ refer to the pion and nucleon field operators, respectively, 
and $\vec \sigma$ ($\fet \tau$) denote the spin (isospin) Pauli matrices.  Further, $g_A$ ($F_\pi$) is the nucleon 
axial coupling (pion decay) constant and $C_{S,T}$ are the LECs
accompanying the leading contact  operators. 
The ellipses refer to terms involving more pion fields. It is
important to emphasize that chiral symmetry leads to highly nontrivial 
relations between the various coupling constants. For example, the strengths of all $\Delta_i=0$-vertices without 
nucleons with $2, 4, 6, \ldots$ pion field operators are given in terms of $F_\pi$ and $M_\pi$. Similarly, 
all single-nucleon $\Delta_i=0$-vertices with $1,2,3,\ldots$ pion fields are expressed in terms of just 
two LECs, namely $g_A$ and $F_\pi$.   We refer the reader
to Refs.~\cite{54959,54960} for further details on the construction of
the effective chiral Lagrangians (for a modern way to construct the pertinent
pion-nucleon Lagrangian, see e.g Ref.~\cite{Fettes:2000gb}).      

The expressions for the power
counting given above are derived under the assumption that there are no
infrared divergences. This assumption is violated for
a certain class of diagrams involving two and more nucleons due to the
appearance of  pinch singularities of the kind
\beq
\label{example}
\int d l_0 \frac{i}{l_0 + i \epsilon} \,  \frac{i}{l_0 - i \epsilon}
\,.
\eeq
Here, $i/(l_0 + i \epsilon)$ is the free nucleon propagator in the
heavy-baryon approach corresponding to the Lagrangian in
Eq.~(\ref{lagr}). 
Clearly, the divergence
is not ``real'' but just an artefact of the extreme non-relativistic approximation
for the propagator which is not applicable in that
case. Keeping the first correction beyond the static limit, the
nucleon propagator takes the form $i/(l_0 - \vec l \,^2/(2m_N) + i
\epsilon)^{-1}$ leading to a finite result for the integral in
Eq.~(\ref{example}) which is, however, enhanced by a factor $m_N / | \vec
q \, |$ as compared to the estimation based on naive dimensional
analysis. In physical terms, the origin of this enhancement is related
to the two-nucleon Green's function of the Schr\"odinger equation
(\ref{SE}). The nuclear potential $V$ we are actually interested in
is, of course,  well defined in the static limit $m_N \to \infty$ and thus
not affected by the above mentioned infrared enhancement. The precise
relation between the nuclear potentials and the amplitude
corresponding to a given Feynman diagram will be discussed in the next
section. 

It is now instructive to address the qualitative implications of the power
counting in Eq.~(\ref{powc_w}) and the explicit form of the effective
chiral Lagrangian.  First, one observes that the dominant contribution
to the nuclear force arises from two-nucleon tree-level diagrams with
the lowest-order
vertices. This implies that the nuclear force is dominated by the
one-pion exchange potential and the two contact
interactions without derivatives. Pion loops are suppressed by two powers of the soft
scale. Also vertices with $\Delta_i >0 $  involving more derivatives   
are suppressed and do not contribute at lowest order. One also
observes the suppression of many-body forces: according to
Eq.~(\ref{powc_w}), $N$-nucleon forces start contributing at order 
$Q^{-4 + 2N}$.  This implies the dominance of the two-nucleon force
with three- and four-nucleon forces  appearing as corrections at
orders $Q^2$ and $Q^4$, respectively.    

\subsection{Derivation of the nuclear forces}  

We now clarify the meaning of the nuclear potential $V$ and
outline some approaches that can be used to derive it. We first rewrite Eq.~(\ref{SE}) as the
Lippmann-Schwinger (LS) equation for the half-shell T-matrix 
\beq
\label{LSE}
T_{\alpha \beta} = V_{\alpha \beta} + \sum_\gamma V_{\alpha \gamma}
\frac{1}{E_\beta  - E_\gamma + i \epsilon}  T_{\gamma \beta}
\eeq
where $\alpha$,  $\beta$ and $\gamma$ denote the few-nucleon
states, $E_\alpha$ is the kinetic energy of the nucleons in the
state $\alpha$  and 
$\sum_\gamma$ is to be understood as a sum (integral) over all
discrete (continuous) quantum numbers of the nucleons.  
On the other hand, the scattering matrix $S$, which is related to
the $T$-matrix via 
\beq
\label{SM}
S_{\alpha \beta} 
= \delta (\alpha - \beta) - 2 \pi i \delta ( E_\alpha - E_\beta ) T_{\alpha \beta}\,,
\eeq
can be directly computed 
from the effective chiral Lagrangian using the Feynman graph
technique. This then allows one to define the potential $V_{\alpha \beta}$ by matching the
amplitude to the iterative solution of Eq.~(\ref{LSE}) which in the
operator form can be written as 
\beq
\hat T = \hat V + \hat V \, \hat G_0 \, \hat V +  \hat V
\, \hat G_0 \, \hat V 
 \, \hat G_0 \, \hat V  + \ldots \,,
\eeq
where $\hat G_0$ is the $A$-nucleon resolvent operator.  The outlined 
approach is, of course, not new and has been extensively used in the fifties of
the last century in the context of the meson field theory, see
e.g.~\cite{Nambu:1950rs,Taketani:1952}. There is, however, one
subtlety here related to the fact that the T-matrix 
calculated from the effective Lagrangian using Feynman diagrams and
entering Eq.~(\ref{SM})  is only available
on the energy shell. On the other hand,  the potential to be substituted in the    
LS equation is needed off the energy shell. This results in an ambiguity
in the definition of the potential corresponding to the freedom in
carrying out off-the-energy-shell extension. This should not come as a
surprise given that the
Hamiltonian $H_0 + V$ is not an observable quantity. 

An alternative method to define nuclear forces exploits another
old idea of decoupling the pion states from the rest of the Fock
space by means of a suitably chosen unitary transformation 
\cite{Okubo:1954zz}. 
This approach was formulated in the context of chiral EFT in Refs.~\cite{Epelbaum:1998ka}.  
The derivation of the unitary operator, nuclear
forces and currents can be carried out straightforwardly using perturbation theory 
in powers of $Q$ and employing the ``algebraic'' version of the power
counting in Eq.~(\ref{pow_co_alg}). The above mentioned ambiguity of the
nuclear potentials and currents can be systematically
explored in this approach by performing further unitary transformations
after decoupling the pion states. Interestingly, this ambiguity turns
out to be strongly constrained by renormalizability of the Hamiltonian
\cite{Epelbaum:2006eu}. 

\begin{figure}[t!]
\includegraphics[width=\textwidth,keepaspectratio,angle=0,clip]{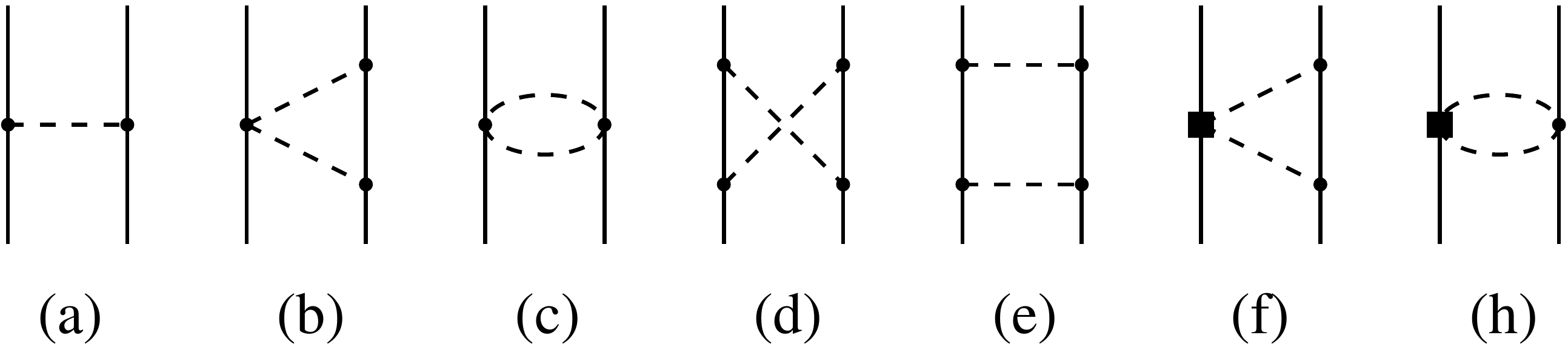}
    \caption{\baselineskip 12pt 
         Diagrams contributing to the long-range part of the
         two-nucleon potential at LO (a), NLO (b)-(e) and N$^2$LO (f), (h). Solid and dashed lines
         represent nucleons and pions, respectively. Solid dots and
         filled rectangles refer to the
         leading  ($\Delta_i = 0$) and subleading ($\Delta_i  = 1$) vertices
         from the chiral Lagrangian.   
\label{fig2} 
 }
\end{figure}

To be specific, consider the derivation of the long-range
two-nucleon potentials  up to next-to-next-to-leading 
order (N$^2$LO). For the sake of simplicity, we use here the matching approach and
follow closely Ref.~\cite{Kaiser:1997mw}, see also Ref.~\cite{Ordonez:1995rz} for a
pioneering calculation within the framework of time-ordered perturbation theory. 
We will not consider here the short-range part of the nuclear force
since it can be directly read off from the Lagrangian. Notice that it is not necessary to
explicitly evaluate pion loop diagrams involving contact interactions
unless one is interested in the quark-mass dependence of the
short-range operators.  As already pointed out before, the
leading-order (LO) contribution $\sim Q^0$ is due to one-pion ($1\pi$)
exchange. 
Evaluating the contribution of diagram (a) in Fig.~\ref{fig2} for on-shell nucleons
yields
\beq
V_{1\pi}^{(0)} = T_{1\pi}^{(0)} \Big|_{E_{\vec p} = E_{\vec p \, '}}=
- \bigg( \frac{g_A}{2 F_\pi} \bigg)^2 \frac{\vec
  \sigma_1 \cdot \vec q\, \vec \sigma_2 \cdot \vec q}{\vec q \, ^2 +
  M_\pi^2} \, \fet \tau_1 \cdot \fet \tau_2\,,
\eeq
where $\vec q = \vec p \, ' - \vec p$ is the nucleon momentum transfer, 
while $\vec p$, $\vec p \, '$ refer to the center-of-mass
(CMS) initial and final momenta.  Notice that while $V_{1\pi}^{(0)}$
is uniquely defined in this procedure in the static limit with
$E_{\vec p} =
E_{\vec p \, '}=0$,  the relativistic corrections are not since one is,
in principle,
free to add to $V_{1\pi}$ terms proportional to $E_{\vec p} -
E_{\vec p \, '}$. 

Because of parity conservation, next-to-leading order (NLO) corrections to the potential
appear at order $Q^2$ rather than $Q$ \cite{UTTG-31-90}. For all
two-pion ($2\pi$) exchange diagrams in Fig.~\ref{fig2} except the box
graph (e), the potential can be defined via the
identification $V_{2\pi}^{(2)} = T_{2\pi}^{(2)} \big|_{E_{\vec p} =
  E_{\vec p \, '}}$. In the case of the box diagram, we have to
subtract the iterated one-pion exchange contribution 
$\hat V_{1\pi}^{(0)} \hat G_0  \hat V_{1\pi}^{(0)} $ to avoid double
counting. Evaluating the corresponding Feynman diagram in the CMS, one obtains
the contribution proportional to the integral
\beqa
\label{matching}
&& \int \frac{d^4 l}{(2 \pi )^4} \frac{(2 m_N)^2 i }{[(p-l)^2 - m_N^2 + i
  \epsilon] [(p+l)^2 - m_N^2 + i
  \epsilon] [l_1^2 - M_\pi^2 + i \epsilon]  [l_2^2 - M_\pi^2 + i
  \epsilon] }  \nn
&&= \int \frac{d^3 l}{(2 \pi )^3} \bigg( 
\frac{1}{\omega_1^2 ( E_{\vec p} - E_{\vec p - \vec l} + i \epsilon )
  \omega_2^2}  + \frac{\omega_1^2 + \omega_1 \omega_2 + \omega_2^2 }{2
\omega_1^3 \omega_2^3 (\omega_1 + \omega_2 )} + \mathcal{O} (m_N^{-1})\bigg)\,,
\eeqa
where $\omega_i = \sqrt{\vec l \, ^2 + M_\pi^2}$ and the virtual pion
momenta are given by $l_1 = l$, $l_2=l+q$.  Notice that we used here the
  relativistic expressions for the nucleon propagators in order to
  avoid the pinch singularity discussed in the previous section.  
The first term in the round brackets is nothing but the iterated one-pion
exchange, $\hat V_{1\pi}^{(0)} \hat G_0  \hat V_{1\pi}^{(0)} $, while
the second one gives rise to $V_{2\pi , \; \rm box}^{(2)}$. Notice
further 
that, as explained above, the iterated $1\pi$ exchange is enhanced
compared to the estimation $\sim Q^2$ based on naive dimensional analysis
since $E \sim \mathcal O (Q^2/m_N) \ll \mathcal{O} (Q)$. 

The final
result for the $2\pi$-exchange potential at order $Q^2$ can now be
obtained by evaluating the loop integrals and carrying out the
spin-isospin algebra. Employing dimensional regularization and using
the decomposition 
\beq
\label{2PEdec}
V = V_C + \fet \tau_1 \cdot \fet \tau_2 \, W_C + \left[   
V_S + \fet \tau_1 \cdot \fet \tau_2 \, W_S \right] \, \vec \sigma_1 \cdot \vec \sigma_2 
+ \left[ V_T + \fet \tau_1 \cdot \fet \tau_2 \, W_T \right] 
\, \vec \sigma_1 \cdot \vec q \, \vec \sigma_2 \cdot \vec q \,,
\eeq
it takes the form 
\beqa
\label{2PE_nlo}
W_C^{(2)} &=& - \frac{L (| \vec q \, | )}{384 \pi^2 F_\pi^4}
\bigg[4M_\pi^2 (5g_A^4 - 4g_A^2 -1)  + \vec q
\, ^2(23g_A^4 - 10g_A^2 -1)
+ \frac{48 g_A^4 M_\pi^4}{4 M_\pi^2 + \vec q \, ^2} \bigg] \,, \nn
V_T^{(2)} &=& -\frac{1}{\vec q \, ^2} V_S^{(2)}  = - \frac{3 g_A^4}{64
  \pi^2 F_\pi^4} \,L ( | \vec q \, |)\,,
\eeqa
where we do not show the contributions that are polynomial in momenta as these 
can be absorbed into the contact interactions. The loop function $L$ is
defined via
\beq
\label{def_LA}
L ( | \vec q \, |) = \frac{\sqrt{4 M_\pi^2 + \vec q\, ^2}}{| \vec q \, |} \ln
\frac{\sqrt{4 M_\pi^2 + \vec q\, ^2} + | \vec q \, |}{2 M_\pi} \,.
\eeq
Notice that the ultraviolet divergences entering the loop integrals
are polynomial in the external momenta and, therefore, do not affect
the non-polynomial pieces if one uses dimensional regularization or
equivalent schemes. 

The next-to-next-to-leading order (N$^2$LO) corrections emerge at
order $Q^3$.  Again, parity conservation forbids any
$1\pi$-exchange contributions at this order. The $2\pi$-exchange terms
emerge from 
triangle diagram (f) in Fig.~\ref{fig2} through the identification
$V_{2\pi}^{(3)} = T_{2\pi}^{(3)} \big|_{E_{\vec p} =   E_{\vec p \, '}}$ 
while graph (h) yields a vanishing result. One finds
\beqa
\label{2PE_nnlo}
V_C^{(3)}  &=& -\frac{3g_A^2}{16\pi F_\pi^4}  \bigg[2M_\pi^2(2c_1 -c_3) -c_3 q^2 \bigg] 
(2M_\pi^2+q^2) A ( | \vec q \, | )\,, \nn
W_T^{(3)} &=& -\frac{1}{q^2} W_S^{(3)}  = - \frac{g_A^2}{32\pi F_\pi^4} \,  c_4 (4M_\pi^2 + q^2) 
A( | \vec q \, | )\,,
\eeqa
where the $c_{i}$ are LECs associated with the $\pi \pi
NN$  vertices of order $\Delta_i=1$  and the loop function $A$ is given by   
\beq
A ( | \vec q \, | ) = \frac{1}{2  | \vec q \, |} \arctan \frac{ | \vec q \, |}{2 M_\pi} \,.
\eeq
In addition to the static terms, there are, in principle, also $1/m_N$-corrections
to NLO graphs, see e.g.~the last term in the brackets in
Eq.~(\ref{matching}). The nucleon mass is, however, often treated as
a very heavy scale with $m_N \gg \Lambda$, see Ref.~\cite{UTTG-31-90} for a
discussion, leading to a suppression of the $1/m_N$ corrections.   

It is now instructive to address the convergence of the chiral
expansion for the long-range two-nucleon force.  Given that the  
obtained expressions depend solely on the momentum transfer $\vec q$,
the potential is expected to be local in coordinate space:  
\beq
V (\vec r \, )= \tilde V_C + \fet \tau_1 \cdot \fet \tau_2 \, \tilde W_C +  \left[ 
\tilde V_S + \fet \tau_1 \cdot \fet \tau_2 \, \tilde W_S \right] \, \vec \sigma_1 \cdot \vec \sigma_2 
+ \left[ \tilde V_T + \fet \tau_1 \cdot \fet \tau_2 \, \tilde W_T \right] S_{12}\,,
\eeq
where $\vec r$ is the distance between the nucleons, $S_{12} \equiv (3 \vec \sigma_1 \cdot \vec r \, \vec \sigma_2
\cdot \vec r  - \vec \sigma_1 \cdot \vec \sigma_2 r^2 ) /r^2$ and
$\tilde V_X$, $\tilde W_X$ are scalar functions of $r\equiv | \vec r 
\, |$.   
The Fourier transform of the expressions in
Eqs.~(\ref{2PE_nlo}) and (\ref{2PE_nnlo}) is, however, ill-defined as the potentials
are not bounded as $q$ increases, where $q \equiv | \vec q \, |$.  At  finite distances $ r > 0$, 
the potential can be obtained through a suitable regularization  
\beq
V ( \vec r \, ) = \lim_{\Lambda \to \infty} \int \frac{d^3 q}{(2 \pi )^3 }\, 
e^{-i \vec q \cdot \vec r} \,
V  (\vec q \, ) \, F_\Lambda \left( | \vec q \, | \right) \,,
\eeq
where the regulator function $F_\Lambda (x)$ can e.g.~be chosen as
$F_\Lambda(x) = \exp (-x^2/\Lambda^2)$. Alternatively and more
elegantly, one can write the functions $W_X$ and $V_X$ 
in terms of a continuous superposition of Yukawa functions 
which
can easily be Fourier transformed, see Ref.~\cite{Kaiser:1997mw} for more details. 
For example, for central potentials one obtains the unsubtracted
dispersive  representation
\beq
\label{disp}
V_C ( q) = \frac{2}{\pi} \int_{2 M_\pi}^\infty
 d \mu \, \mu \frac{\rho_C (\mu )}{\mu^2 + q^2} \,, \quad \quad
V_C ( r) = \frac{1}{2 \pi^2 r} \int_{2 M_\pi}^\infty d\mu \, \mu
e^{-\mu r} \rho_C (\mu )\,,
\eeq 
where $\rho_C (\mu ) = {\rm Im}~[V_C (0^+ - i \mu)]$ is the
corresponding spectral function.  
 
\begin{figure}[t!]
\centerline{\includegraphics[width=\textwidth,keepaspectratio,angle=0,clip]{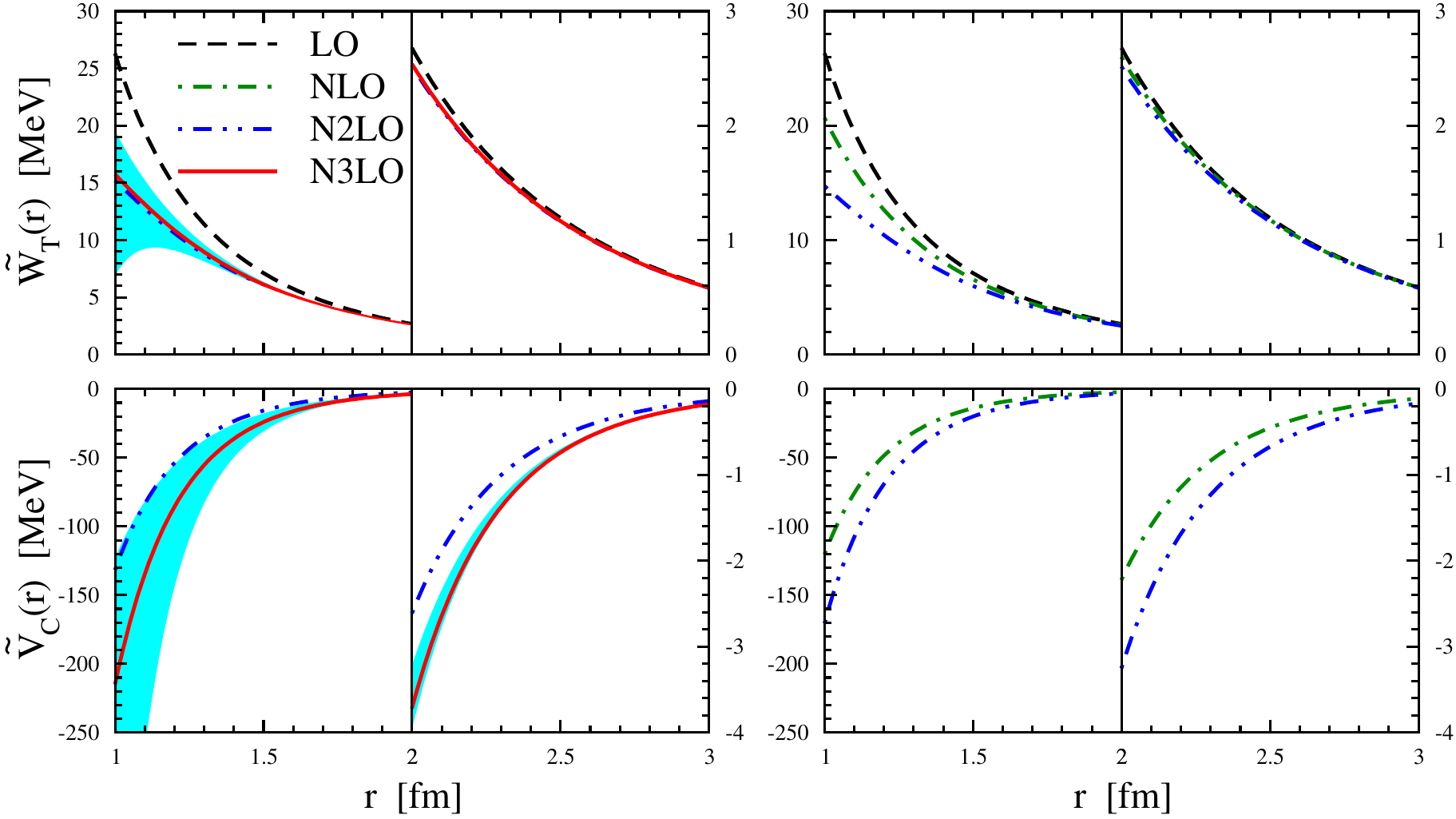}}
    \caption{\baselineskip 12pt
         Chiral expansion of the isovector-tensor (upper row) and
         isoscalar central (lower row) long-range potentials $\tilde
         W_T (r)$ and $\tilde V_C (r)$, respectively.  The left (right)
         panel shows the results for the EFT without (with) explicit
         $\Delta$(1232) degrees of freedom. The light-shaded band shows
         the estimation of the intrinsic model dependence associated
         with the short-range components as explained in the text (only
         shown for the theory without deltas). 
\label{fig3} 
 }
\end{figure}

In Fig.~\ref{fig3} we show the chiral expansion for the two most
important cases, namely for the 
isovector-tensor and isoscalar-central potentials $\tilde W_T (r)$ and $\tilde
V_C (r)$.  We also include the contributions at next-to-next-to-next-to-leading
order (N$^3$LO) whose explicit form can be found in
Ref.~\cite{Kaiser:1999ff}
but restrict ourselves to the local pieces omitting the $1/m_N$ corrections. The shaded bands in the
figure visualize the estimated scheme dependence which is intrinsic to 
the separation between the long- and short-range contributions in the
potential. Specifically, we only include in the dispersive integrals in
Eq.~(\ref{disp}) the components in the spectrum with $\mu < \tilde
\Lambda = 1\,$GeV.  
The high-$\mu$ components generate terms which, at low momenta, 
are indistinguishable from contact interactions parameterizing the
short-range part of the chiral potential. The bands correspond to the
variation of $\tilde \Lambda$ in the range from  $800\,$MeV to  $\infty$. Their
widths may, therefore, serve as an estimation of the size of
short-range components not associated with the dynamics of Goldstone
Bosons. 

The potential in the
isovector-tensor channel is clearly dominated by one-pion exchange $V_{1\pi}$. 
Two-pion exchange contributions  in this channel become visible
at distances of the order $r \sim 2$~fm and smaller. The strong, 
attractive isoscalar-central potential of intermediate range is another well-known
feature of the two-nucleon force. Phenomenologically, it is attributed
to the correlated two-pion exchange which is often modeled in terms
of the $\sigma$-meson exchange \cite{Machleidt:1989tm}. In chiral EFT,
on the other hand, 
all low-energy manifestations of the $\sigma$ and other
heavy mesons are systematically taken into account through values of
the LECs in the effective Lagrangian.    
The resulting strength of $\tilde V_C$ turns out to be comparable to that of
$V_{1\pi}$ even at distances $r \sim 2$~fm and appears to be an order of magnitude 
bigger than the strength of two-pion exchange in any other channel. 
The large size of the N$^2$LO contributions can be traced back to the
large numerical factor of $3/16$ in the case of $V_C^{(3)}$,   
an enhancement by one power of 
$\pi$, cf.~Eqs.~(\ref{2PE_nlo}) and (\ref{2PE_nnlo}), and the large
values of the LECs $c_{3,4}$. For example, the determination from
pion-nucleon subthreshold coefficients at order $Q^2$ leads to 
$c_3 = -3.9$~GeV$^{-1}$ and $c_4 = 2.9$~GeV$^{-1}$ 
\cite{Bernard:1996gq,Epelbaum:2007sq} which are larger in
magnitude than the expected natural size $| c_{3,4} | \sim g_A/\Lambda \sim 
1\,$GeV$^{-1}$.  Moreover, even larger values are obtained from
pion-nucleon scattering at order $Q^3$ where
effects of pion loops are taken into account \cite{Fettes:1998ud,Buettiker:1999ap}.
The large values of $c_{3,4}$ can be traced
back to the implicit treatment of the $\Delta$(1232) isobar. Given the 
fairly low excitation energy of the Delta $m_\Delta - m_N$, which is
numerically $\sim 2 M_\pi$, one may expect that its explicit inclusion
in the EFT within the so-called small scale expansion
\cite{Hemmert:1997ye} based on the
(phenomenological) extension of the counting in Eq.~(\ref{deltaless}) to 
\beq
\label{deltafull}
\varepsilon \in \bigg\{ \frac{M_\pi}{\Lambda}, \,    \frac{| \vec k\, |
}{\Lambda} , \, \frac{m_\Delta - m_N}{\Lambda} \bigg\}\,,
\eeq
will allow one to resum a certain class of important contributions
leading to a superior convergence as compared to the delta-less
theory. The improved convergence is indeed observed both in the case
of pion-nucleon scattering \cite{Fettes:2000bb} and nuclear forces 
\cite{Kaiser:1998wa ,Krebs:2007rh,Epelbaum:2007sq}.  In particular, 
the dominant contribution to $V_C$ and $W_T$ emerges in the
$\Delta$-full theory already at NLO with the N$^2$LO contributions
providing fairly small corrections, see the right panel in
Fig.~\ref{fig3}. For example, the single $\Delta$-excitation in the 
box diagrams (d) and (e) of Fig.~\ref{fig2} generates at NLO the
isoscalar-central potential 
\beq
 V_C^{(2)}  = -\frac{g_A^2 h_A^2}{12\pi F_\pi^4 (m_\Delta - m_N)}  
(2M_\pi^2+q^2)^2 A ( | \vec q \, | )\,,
\eeq
where $h_A$ denotes the $\pi N \Delta$ axial coupling. In the standard
$\Delta$-less approach based on the assignment $m_\Delta - m_N \sim
\Lambda \gg M_\pi$, this numerically large contribution is shifted to
N$^2$LO where it is reproduced
through the $\Delta$-isobar saturation of $c_3$, $c_3^\Delta = -
4 h_A^2/(9 (m_\Delta - m_N))$ \cite{Bernard:1996gq}, see
Eq.~(\ref{2PE_nnlo}). 
Having
included the effects of the $\Delta$-isobar explicitly, one finds 
strongly reduced values of the LECs $c_{3,4}$ in agreement with the
naturalness assumption. For example, using
$h_A= 3 g_A/(2 \sqrt{2})$ 
from SU(4) or large $N_c$, one obtains 
$c_3 = -0.8$ GeV$^{-1}$ and $c_4 = 1.3$ GeV$^{-1}$ \cite{Krebs:2007rh}.
Thus, the major part of the unnaturally large subleading
$2\pi$-exchange potential at N$^2$LO is shifted to NLO in the
$\Delta$-full theory. This more natural convergence pattern 
is visualized in the right panel of Fig.~\ref{fig3}.  

When substituted into  the Schr\"odinger equation, the long-range
potentials introduced above provide an approximate representation of
the nearby left-hand singularities in the partial wave amplitude  as shown in
Fig.~\ref{fig1}. These cause a rapid energy dependence and imply non-trivial
relations between the coefficients in the effective-range expansion 
which can be regarded as low-energy theorems (LETs)
\cite{Cohen:1998jr,Cohen:1999iaa,Epelbaum:2009sd} and confronted with
the data. A pedagogical introduction 
to the LETs and their relation to the
so-called modified effective range expansion can be found in
Ref.~\cite{Epelbaum:2010nr}.  Assuming perturbativeness of the  
pion-exchange contributions, these relations can be worked out
analytically within the scheme proposed by Kaplan, Savage and Wise in
Ref.~\cite{Kaplan:1998tg}. The resulting LETs, however, appear to
be strongly violated in the $^1S_0$ and the $^3S_1$-$^3D_1$ channels
\cite{Cohen:1998jr,Cohen:1999iaa}. This indicates the non-perturbative
nature of the one-pion
exchange potential at least in these channels, see
also Ref.~\cite{Fleming:1999ee}. Employing a non-perturbative treatment of 
the pion-exchange potential, the LETs were tested numerically in
Ref.~\cite{Epelbaum:2004fk}. A closely related approach is followed by
Birse et al.~in Refs.~\cite{Birse:2003nz,Birse:2007sx,Birse:2010jr,Ipson:2010ah} by
analyzing the  energy dependence of the residual short-range potential in a
given partial wave.  
Perhaps the most impressive evidence of the chiral $2\pi$-exchange
comes from the Nijmegen partial wave analysis of
proton-proton scattering \cite{Rentmeester:1999vw}. Here, the
Schr\"odinger equation is solved for a specific choice of the long-range
potential outside of some boundary $b$. Short-range physics is then taken
into account by choosing appropriate boundary conditions at $r=b$. The
number of parameters needed to describe experimental data below the pion
production threshold with $\chi^2_{\rm datum} \sim 1$ can be regarded
as a measure of the
amount of physics not included in the long-range potential. With
$b=1.4$ fm, the authors of Ref.~\cite{Rentmeester:1999vw} observe a
reduction $31 \to 28 \to 23$ in the number of parameters when
employing $V_{1\pi} \to V_{1\pi} + V_{2\pi}^{(2)} \to V_{1\pi} +
V_{2\pi}^{(2)} + V_{2\pi}^{(3)}$ as the long-range potential
(in addition to the corresponding electromagnetic interactions).

\subsection{Nuclear forces: the status and open issues} 
\label{sec_forces}

\begin{figure}[t!]
\centerline{\includegraphics[width=\textwidth,keepaspectratio,angle=0,clip]{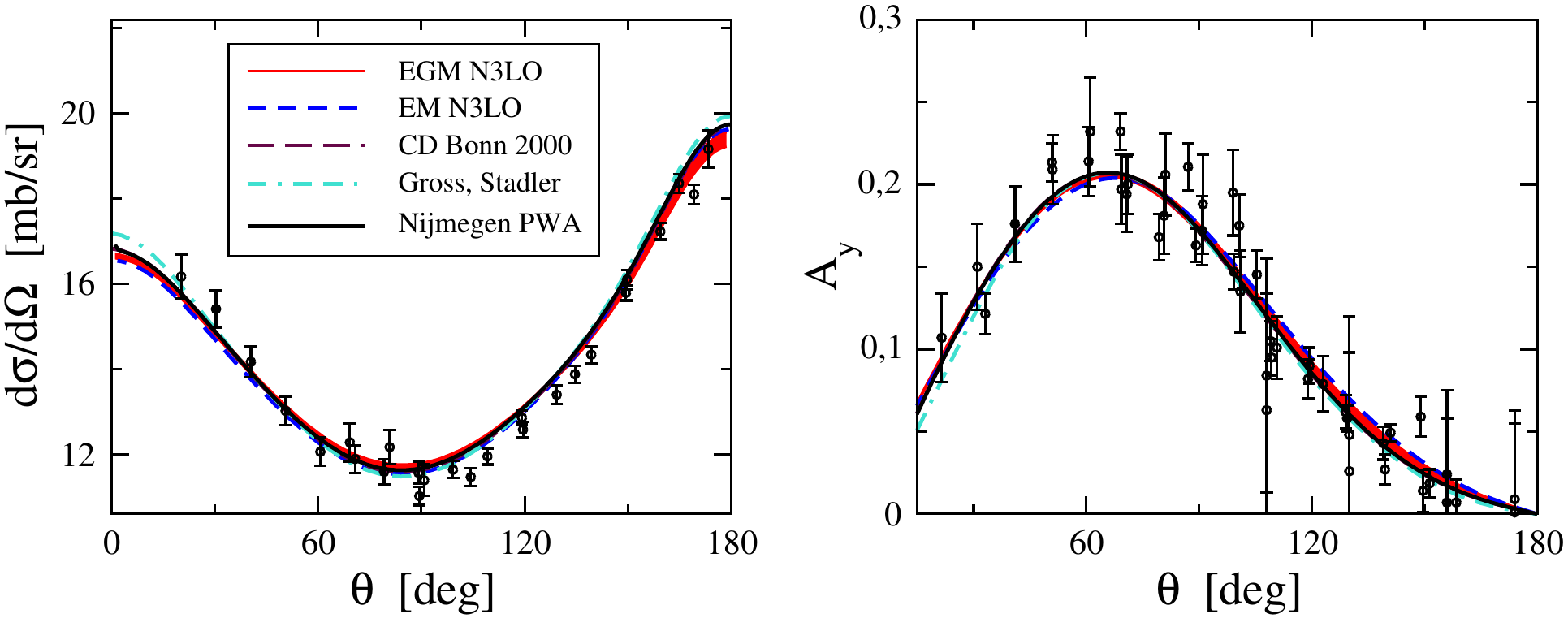}}
    \caption{\baselineskip 12pt
         Neutron-proton differential cross section (left panel) and
         analyzing power (right panel) at $E_{\rm lab} = 50$ MeV
         calculated using chiral EFT, the CD Bonn 2000 potential of
         Ref.~\cite{Machleidt:2000ge} and the potential developed by Gross and Stadler
         in Ref.~\cite{Gross:2008ps}. Also shown are results
         from the Nijmegen partial wave analysis \cite{Stoks:1993tb}. 
         References to data can be found in  \cite{Stoks:1993tb}. 
\label{fig4} 
 }
\end{figure}

We now summarize the current status of the nuclear forces within the 
heavy-baryon, $\Delta$-less formulation based on the power counting of
Eq.~(\ref{powc_w}). 
In this scheme, the nuclear Hamiltonian is presently worked out up to
N$^3$LO in the chiral expansion,
\beq
H = H_0 + V_{\rm 2N} +  V_{\rm 3N} +  V_{\rm 4N} + \ldots
\eeq
with 
\beqa
V_{\rm 2N} &=& V_{\rm 2N}^{(0)} + V_{\rm 2N}^{(2)} + V_{\rm 2N}^{(3)}
+ V_{\rm 2N}^{(4)} + \ldots \,, \nn
V_{\rm 3N} &=& V_{\rm 3N}^{(3)} + V_{\rm 3N}^{(4)}  + \ldots \,, \nn
V_{\rm 4N} &=& V_{\rm 4N}^{(4)}  + \ldots \,,
\eeqa
where the ellipses refer to terms beyond N$^3$LO.  
For two nucleons, it turns out to be necessary and
sufficient to go to N$^3$LO  in order to accurately describe  
the neutron-proton and proton-proton phase shifts up to laboratory
energies of  $E_{\rm lab} \sim 200$ MeV
\cite{Epelbaum:2004fk,Entem:2003ft} . 
This is visualized in Fig.~\ref{fig4}
where, as a representative example, the experimental data for the
neutron-proton differential cross section and vector analyzing power at $E_{\rm lab}=
 50$ MeV  are shown  in comparison with the calculations based on the chiral NN
potentials of Refs.~\cite{Epelbaum:2004fk,Entem:2003ft} and various
modern  phenomenological potentials.
At the accuracy of N$^3$LO, it is mandatory to take into account 
isospin-breaking (IB) contributions. The dominant IB effects emerge from the
charge-to-neutral pion mass difference in the $1\pi$ and $2\pi$
exchange \cite{Friar:1999zr}  
(charge-independence breaking), proton-to-neutron mass
difference in the $2\pi$-exchange \cite{Friar:2004ca,Epelbaum:2005fd}  
(charge-symmetry breaking) and the
two derivative-less contact interactions (both charge-independence and
charge-symmetry breaking in the two S-waves). The short-range part of
the potential at  N$^3$LO receives contributions from 24
isospin-invariant and 2 IB contact interactions whose strength was
adjusted to phase shifts (scattering data)  in
Ref.~\cite{Epelbaum:2004fk} (Ref.~\cite{Entem:2003ft}).   Both
available versions of the N$^3$LO potential employ a finite
momentum-space cutoff in order to regularize the Schr\"odinger
equation. This cutoff is varied in Ref.~\cite{Epelbaum:2004fk} in the
range $\Lambda = 450\ldots 600$ MeV.  More details on the
construction of chiral potentials at N$^3$LO can be found in the comprehensive
review articles \cite{Epelbaum:2005pn,Machleidt:2011zz}.  

While the chiral expansion of the long-range nuclear forces emerges
rather straightforwardly, the power
counting for the short-range operators and the closely related issue
of non-perturbative renormalization of the Schr\"odinger (or LS) equation 
are still being debated in the community, see 
e.g.~\cite{Nogga:2005hy,PavonValderrama:2005wv,Epelbaum:2006pt}.  
Here, the main conceptual difficulty is associated with  the
non-perturbative treatment of the $1\pi$-exchange
potential $V_{1\pi}$. In spite of a considerable effort, see
e.g.~Refs.~\cite{Kaplan:1998tg,Beane:2008bt}, no approximation to
$V_{1\pi}$ is presently known which would capture the relevant
non-perturbative physics and, at the same time, be  analytically
resummable and renormalizable. One is, therefore, left with the
numerical solution of the Schr\"odinger equation for appropriately
regularized chiral potentials along the lines of
Ref.~\cite{Lepage:1997cs}. This paper also explains the meaning of
renormalization in such an approach and provides a tool to verify 
its consistency a posteriori by means of the so-called Lepage-plots.
Notice that iterating the potential, truncated at a given
order of the chiral expansion, in the LS equation necessarily generates higher-order
contributions in the amplitude, which are, generally,
ultraviolet-divergent and whose renormalization requires counter terms
beyond the truncated potential. It is, therefore, not legitimate in
such an approach to arbitrarily increase the cutoff $\Lambda$. 
This point is exemplified in Ref.~\cite{Epelbaum:2009sd} using an
exactly solvable analytical model,  see also Ref.~\cite{Lepage:1997cs}
for a qualitative discussion.  More work is needed in order to (better)
understand the power counting \emph{for the  
scattering amplitude} in the presence of the long-range pion-exchange 
potentials. A promising tool to address this question is provided by the
modified effective range expansion, see the discussion in Ref.~\cite{Epelbaum:2009zz}
and Refs.~\cite{Birse:2007sx,Birse:2010jr} for related work. Also, it remains to be seen
whether renormalization-group based approach along the lines of
Ref.~\cite{Birse:2010fj} can shed new light on this issue. 

Three-nucleon forces  (3NFs) are an old but still relevant topic in nuclear
physics. In spite of many decades of effort, the detailed structure
of the 3NF is not captured by modern phenomenological 3NF
models. Indeed, the global analysis presented in
Ref.~\cite{KalantarNayestanaki:2011wz} demonstrates that 
the available models do not allow to significantly reduce the observed
discrepancies between the experimental data and calculations based on the
high-precision NN potentials for breakup and polarization
observables in elastic nucleon-deuteron scattering. Given the very rich spin-momentum
structure of the 3NF as compared to the NN force, scarcer database and
relatively high computational cost, further progress in this fields
clearly requires input from theory. This provides a strong motivation
to study the structure of the 3NF within chiral EFT. 

\begin{figure}[t!]
\centerline{\includegraphics[width=\textwidth,keepaspectratio,angle=0,clip]{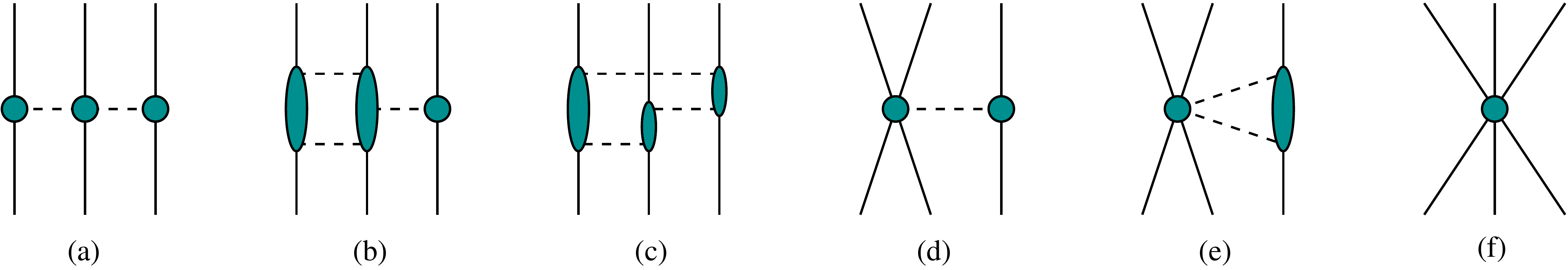}}
    \caption{\baselineskip 14pt
         Various topologies contributing to the 3NF up to order
         $Q^5$. Shaded blobs represent the corresponding amplitudes. 
\label{fig5} 
 }
\end{figure}

The general structure of the 3NF up to order $Q^4$, which also holds
at order $Q^5$, is represented by
six topologies shown in Fig.~\ref{fig5}.  
The first non-vanishing contributions emerge at
N$^2$LO ($Q^3$) \cite{vanKolck:1994yi,Epelbaum:2002vt} from 
the $2\pi$ (a), $1\pi$-contact (d) and contact (f) diagrams. The
corresponding $\pi N$ ($\pi \pi N$, $\pi NN$, $NNN$) amplitudes at
this order are simply given by the $\Delta_i=0$ ($\Delta_i = 1$) vertices
from the effective Lagrangian.  The $1\pi$-contact and contact
diagrams depend on two LECs $c_D$ and $c_E$, respectively, 
whose determination requires few-nucleon
data. We will discuss the applications of the resulting nuclear Hamiltonian to
the properties of few-nucleon systems in the next section. 
It should, however, be emphasized that the leading 3NF involves a rather
restricted set of isospin-spin-momentum structures, which are also
included in the phenomenological 3NF models. In particular, the
longest-range, two-pion exchange topology (a) is well established as one of
the most important phenomenological 3NF mechanisms.  
The leading chiral 3NF can, therefore,
not be expected to shed new light on the persisting
deficiencies in the theoretical description of e.g.~nucleon-deuteron scattering. 
The corrections to the 3NF at N$^3$LO generated by the leading
loop diagrams are worked out in
Refs.~\cite{Ishikawa:2007zz,Bernard:2007sp,Bernard:2011zr}. It is
remarkable that the N$^3$LO terms do not involve any unknown
LECs. In addition to
the static loop contributions, one also has to take into account the 
$1/m_N$ corrections in the topologies (a) and (d)
\cite{Bernard:2011zr}, see also an earlier calculation in
Ref.~\cite{Friar:1994zz}. The numerical implementation of the novel
N$^3$LO contributions in few-body studies is non-trivial and requires
partial wave decomposition. This work is presently in progress, see
Ref.~\cite{Skibinski:2011vi} for a first step in this direction.  

One of the most interesting features of the 3NF at N$^3$LO is,
clearly, its rather rich spin-momentum structure yielding many operators that have never been
explored in few-body studies and might potentially be capable of
resolving the observed discrepancies in 3N scattering.  This
especially applies to the ring topology (c) and, to a lesser extent,
also to the $2\pi$-$1\pi$ topology (b). However, the observed convergence
pattern of the chiral expansion of the $2\pi$-exchange 2N potential with the
leading graphs (b)-(e) in Fig.~\ref{fig2} yielding small contributions and the major
effect emerging from the subleading diagram (f), see the discussion in the
previous section, puts a question mark on the convergence of the 3NF 
at N$^3$LO. Indeed,  
since the (large) LECs $c_{2,3,4}$ saturated by the $\Delta$-isobar do not
contribute to the ring and $2\pi$-$1\pi$ exchange 3NF topologies 
at N$^3$LO, one may expect that the corresponding potentials of
Ref.~\cite{Bernard:2007sp} are not yet converged, see also the discussion in
Ref.~\cite{Machleidt:2010kb}. Thus, one may either need to go to at least N$^4$LO in
the $\Delta$-less theory or to explicitly take into account the contributions of
the $\Delta$-isobar upto  N$^3$LO. This work is in progress, see
Refs.~\cite{Krebs:2011zz} for the first step along these lines.    

The parameter-free results (being) obtained in chiral EFT for the various components
of the 3NF at large distances rely solely on the spontaneously broken chiral
symmetry of QCD. This opens a very interesting possibility for
benchmarking with future lattice QCD calculations,\footnote{\baselineskip 12pt
  Clearly, care is
  required to deal with non-uniqueness of the nuclear
  potentials. The long-range part of the 3NF at N$^3$LO is, however, uniquely
  determined after fixing the corresponding long-range part of the 2NF.} 
 see Refs.~\cite{Ishii:2006ec,Doi:2011nm}
for first attempts along these lines.

Finally, the four-nucleon force (4NF) also receives its first
contribution at N$^3$LO emerging from tree-level diagrams constructed
from the lowest-order vertices of dimension $\Delta_i=0$. The
parameter-free expressions for the 4NF at N$^3$LO can be found in
Ref.~\cite{Epelbaum:2006eu}. The contribution of the 4NF to the
$\alpha$-particle binding energy was estimated in Ref.~\cite{Rozpedzik:2006yi}
to be of the order of a few 100 keV. This provides some justification
for neglecting four- and more-nucleon forces in nuclear structure
calculations.

\section{Applications to few-nucleon systems}

Having determined most of the parameters in the nuclear Hamiltonian 
from nucleon-nucleon data, it is now interesting to test it in
few-nucleon reactions where the $A$-nucleon Schr\"odinger equation
(\ref{SE}) can be solved numerically exactly. For three particles, the
Schr\"odinger equation is conveniently rewritten in terms of the Faddeev
integral equations which are usually solved in the partial wave basis, see
Ref.~\cite{Gloeckle:1995jg}  for details. The Faddeev equations can
nowadays be routinely solved for any given two- and three-nucleon potentials
for both bound and scattering states. For a review of recent progress towards
including the Coulomb interaction in 3N scattering see Ref.~\cite{Deltuva:2010nt}. 

\begin{figure}[t!]
\parbox{7.3cm}{\includegraphics[width=7.3cm,keepaspectratio,angle=0,clip]{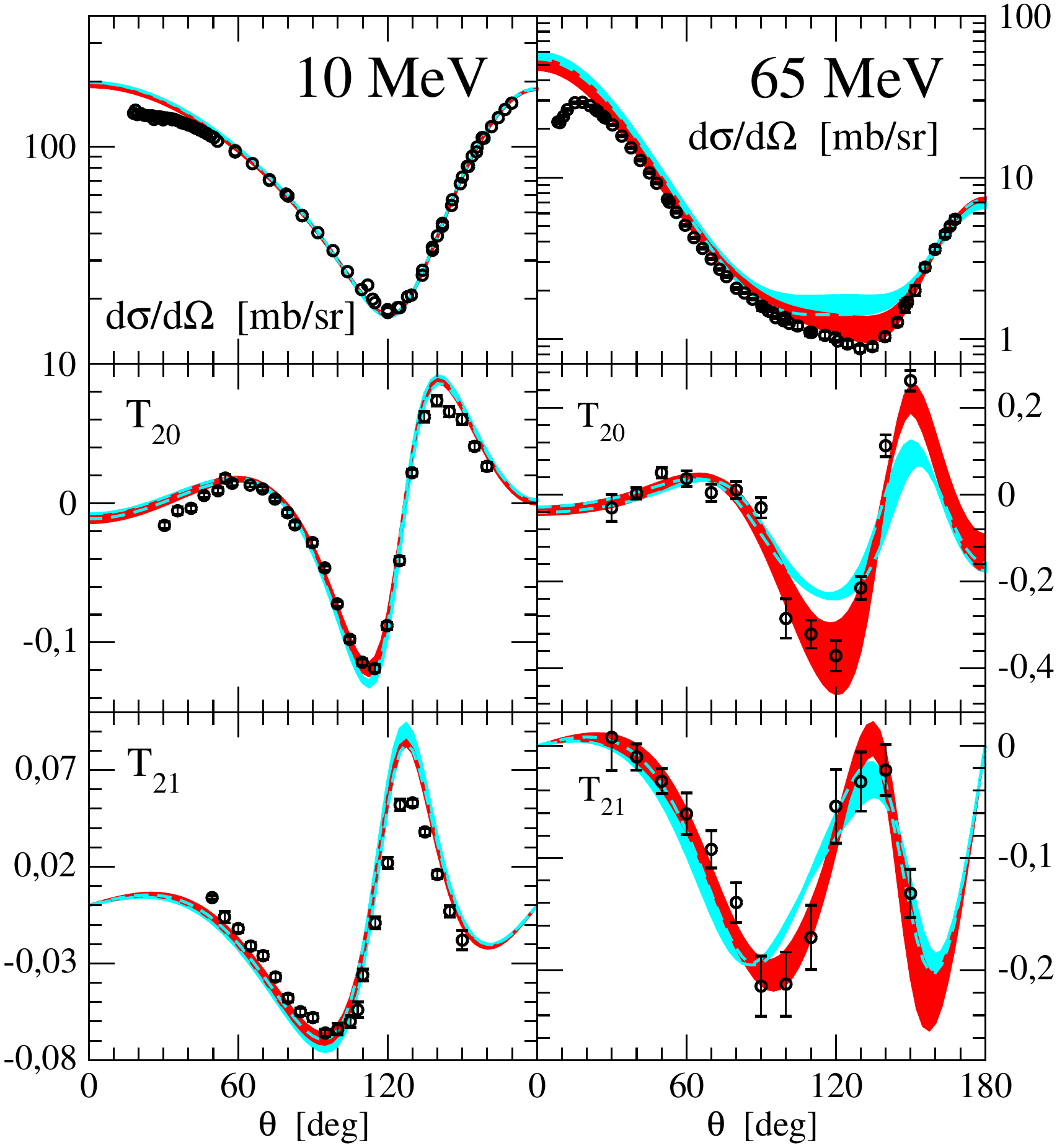}}
\hfill
\parbox{5.84cm}{\includegraphics[width=5.84cm,keepaspectratio,angle=0,clip]{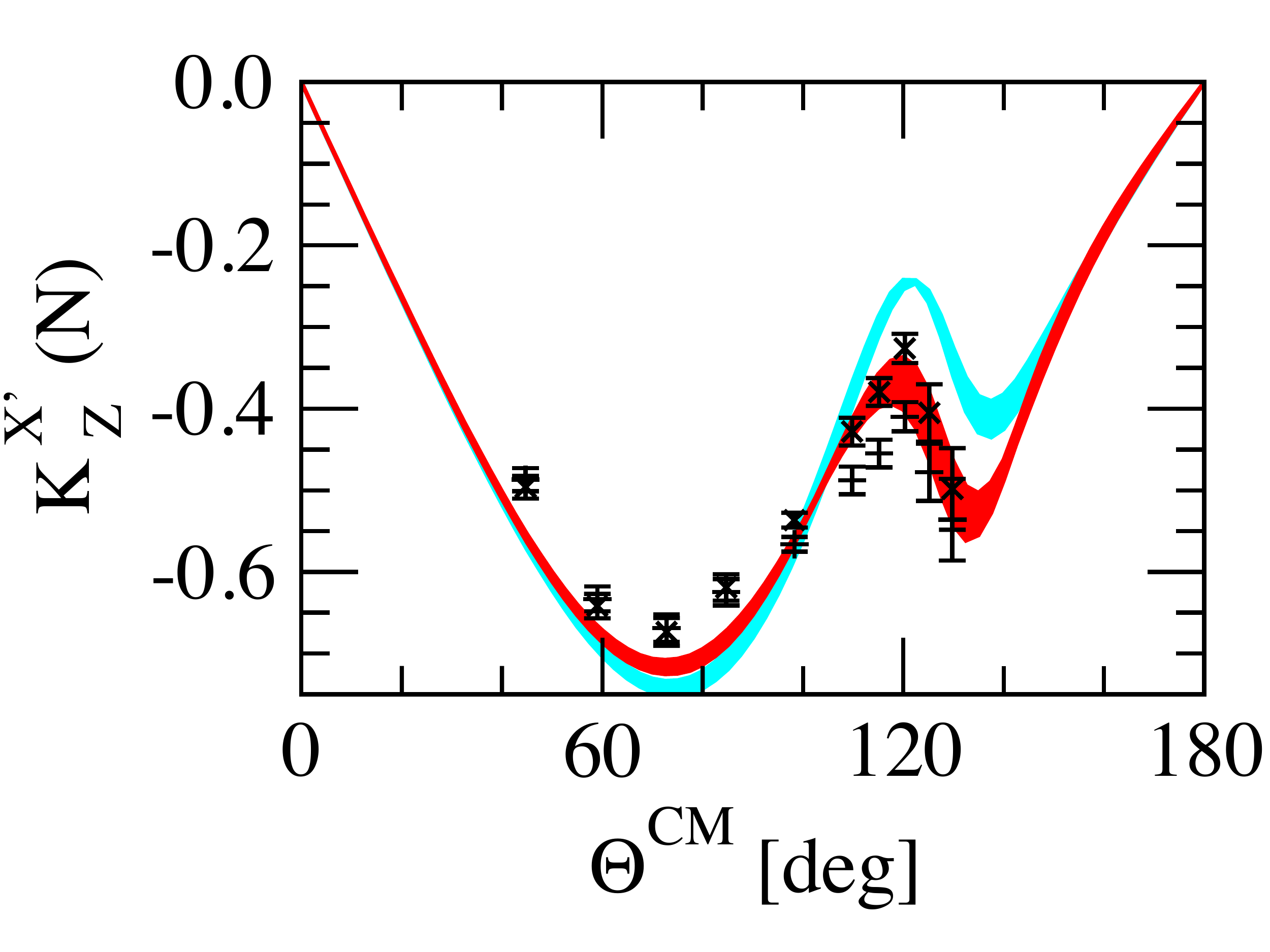}

\hskip 0.13 true cm
\includegraphics[width=5.5cm,keepaspectratio,angle=0,clip]{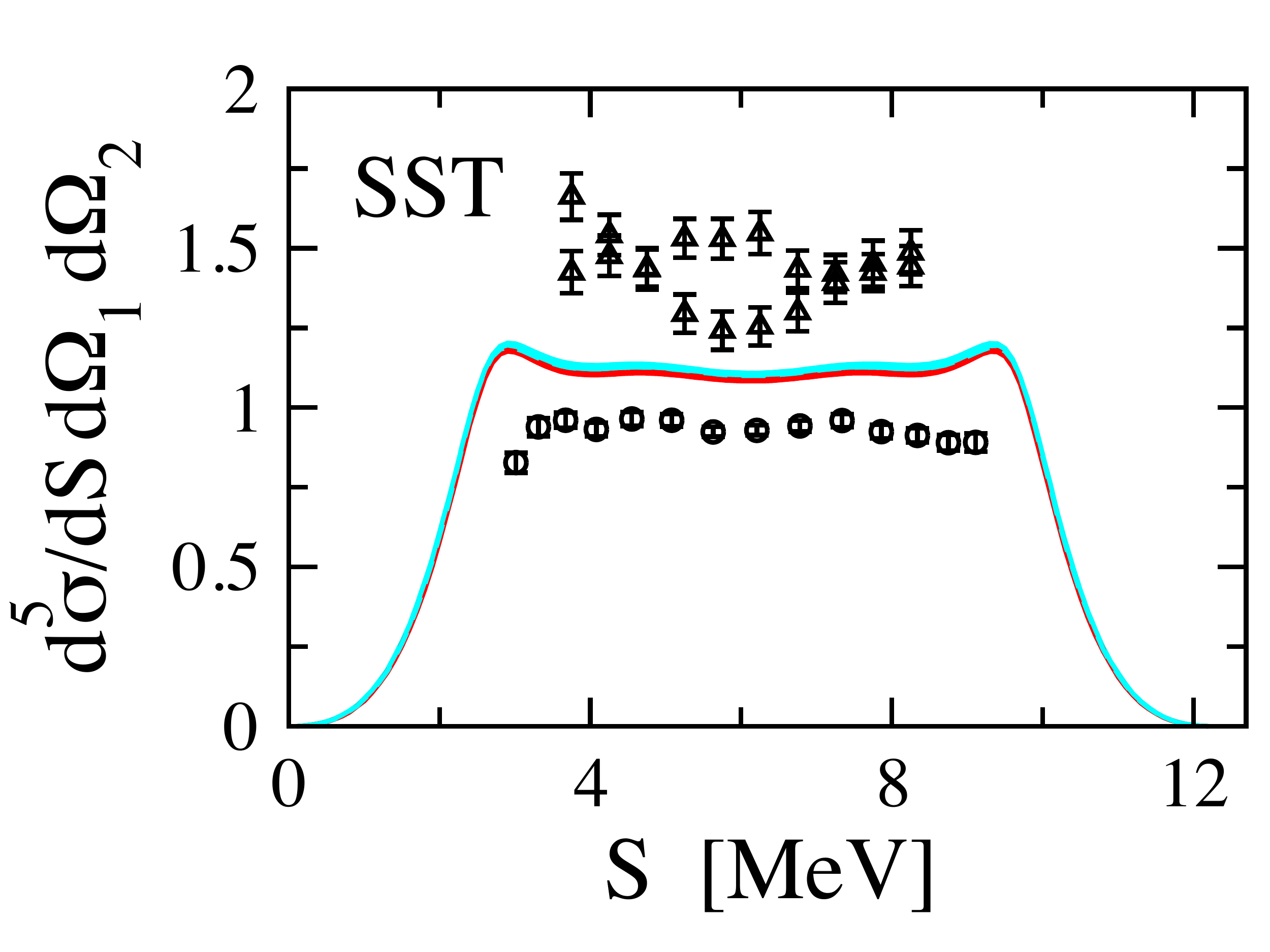}}
    \caption{\baselineskip 12pt
         Left panel: Differential cross section and tensor analyzing
         powers $T_{20}$ and $T_{21}$ for elastic $Nd$ scattering at
         $E_{\rm lab}^N=10$ and $65$ MeV. Right panel, top: The
         nucleon-to-nucleon polarization transfer coefficient in elastic $Nd$
         scattering at $E_{\rm lab}^N=22.7$ MeV (proton-deuteron ($pd$) data from
         Ref.~\cite{kretschmer}). Right panel, bottom: 
         $Nd$ breakup cross section in the space star configuration
         (upper data $nd$, lower data $pd$). Light (blue) and dark (red) shaded bands
         show the results from the chiral EFT at NLO and N$^2$LO, in order. The precise kinematical
         description and references to data can be found in Ref.~\cite{Gloeckle:1995jg}. 
\label{fig6} 
 }
\end{figure}

As explained in the previous section, the 3NF at N$^2$LO depends on
two LECs $c_D$ and $c_E$ which need to be determined from few-nucleon
data. In Ref.~\cite{Epelbaum:2002vt},  $c_D$ and $c_E$ were tuned to
the triton binding energy and the neutron-deuteron ($nd$) doublet scattering
length. The resulting parameter-free nuclear Hamiltonian can then be
tested in nucleon-deuteron ($Nd$) scattering. In Fig.~\ref{fig6}, a
sample of results is shown in comparison with the data. The bands
emerge from the cutoff variation as discussed in the previous
section. Notice that the NLO results shown by the light-shaded
bands are based solely on the two-nucleon force. 
We further emphasize that the calculations shown do not
include the Coulomb interaction and thus correspond to  neutron-deuteron
scattering. We have corrected the proton-deuteron data at
$E_{\rm lab}^N=10$ MeV by subtracting out the (estimated) Coulomb-force
contribution, see Ref.~\cite{Epelbaum:2002vt} for more details. 
Remarkably, even some rather
accurate data for double-polarization observables are
available at low energy. As a representative example, we show in the
top right panel of Fig.~\ref{fig6} our results for the
nucleon-to-nucleon polarization transfer coefficient $K^{x'}_z (N)$ at $E_{\rm
lab}^N = 22.7$ MeV in comparison with the
data of Ref.~\cite{kretschmer}. More results for different
polarization transfer coefficients at this energy, also based on
conventional nuclear potentials, can be found in Ref.~\cite{Witala:2006nn}. 
For most of the elastic observables,
one obtains an improved description when going from NLO to N$^2$LO
which is consistent with the description of two-nucleon data at these
orders. The increasing theoretical uncertainty, however, 
limits the applicability of the N$^2$LO chiral forces to energies below 
$E_{\rm lab}^N \sim 100$ MeV, where the modern phenomenological 2NFs and 3NFs also
provide  an accurate description of the data. There is one well-known
exception from the generally good agreement between the theory and the
data at very low energies known as the $A_y$-puzzle. The puzzle refers
to the strong underprediction of the nucleon vector analyzing power
observed at
energies below $E_{\rm lab}^N \sim 30$ MeV for all modern two- and
three-nucleon potentials, see Ref.~\cite{Gloeckle:1995jg} for more
details. It should, however, be emphasized that $A_y$ is (i) very
small at these energies and (ii) is very sensitive to small
contributions to the nuclear force \cite{Gloeckle:1995jg}, see
Ref.~\cite{KalantarNayestanaki:2011wz} for an extensive discussion. 
It is, therefore, not surprising that the solution to the $A_y$-puzzle
in chiral EFT is not achieved at N$^2$LO and requires the inclusion of
higher-order terms in the Hamiltonian.   

The kinematically very rich deuteron breakup reactions provide even
more detailed insights into nuclear dynamics.
At low energies, only very few selected observables, mainly  the cross
section, are available in certain regions of the phase space. While
a good agreement between the data and the calculations based on
the conventional potentials and chiral EFT is observed for the 
final-state-interaction and quasi-free-scattering configurations,
large discrepancies occur in the case of the space-star configuration 
(the plane in the CMS spanned by  the outgoing nucleons is
perpendicular to the beam axis, and the angle between the nucleons are
$120^\circ$). This is visualized in the right bottom panel of
Fig.~\ref{fig6}, where a comparison should be taken with the upper
sets of neutron-deuteron (nd) data. It is remarkable that the existing 3NFs have
almost no effect in this observable.  For recent studies of related 
breakup configurations see Refs.~\cite{Duweke:2004xv,Ley:2006hu}. 
At higher energies the situation is similar to the elastic channels
with the predictions from chiral EFT being, generally, in agreement
with the data but showing a rapidly increasing theoretical
uncertainty. We refer the readers to Refs.~\cite{Kistryn:2005fi,Stephan:2007zza,Stephan:2010zz}, where the
high-precision cross section and analyzing powers measured recently at
KVI at $E_{\rm lab}^N=65$ MeV and covering  a large part of the available
phase space are confronted with the theoretical calculations. For a
detailed review on 3N scattering at intermediate energies the reader
is referred to \cite{KalantarNayestanaki:2011wz}.  

The four-nucleon (4N) continuum provides another interesting and, given the
appearance of low-energy resonance structures, also very sensitive
testing ground for nuclear dynamics. It also offers the possibility
to probe isospin channels not accessible in nucleon-deuteron
scattering. The solution of the Schr\"odinger equation for
4N scattering states still represents a major challenge so that
only a restricted set of calculations, typically at low energies, is
available. Interestingly, the $A_y$ puzzle persists in the 4N system
where it becomes even more striking due to a much larger magnitude
of $A_y$. The very recent study by the Pisa group
\cite{Viviani:2010mf}  shows 
that, differently to the 3N system, the $A_y$-puzzle in the 4N system
is significantly reduced by the chiral 3NF at N$^2$LO with the LECs
$c_D$ and $c_E$ being adjusted to the $^3$H and $^4$He binding
energies. Given the lack of space, we refrain from a more detailed
discussion of 4N scattering and refer the reader to the review article 
\cite{Viviani:2010mf}.

The nuclear Hamiltonian at N$^2$LO was also applied to compute the
spectra of light nuclei. In particular, one obtains $7.7\ldots 8.5$ MeV
and $24.4 \ldots 28.8$ MeV  for the triton
and $\alpha$-particle binding energies (BE) at NLO. These results agree well
with the experimental values of $8.482$ MeV and $28.30$ MeV,
respectively. With the triton BE being used as input in the
determination of $c_{D,E}$, the $\alpha$-particle
BE at N$^2$LO, $27.8\ldots 28.6$ MeV, is improved as compared to NLO.  

In all applications discussed so far the LECs entering the 3NF were
determined from the triton BE and the neutron-deuteron scattering
length. Given the strong correlation between these observables known
as the Philips line and caused
by the large S-wave scattering lengths in the two-nucleon system, 
the resulting values for $c_{D,E}$ suffer from a sizable uncertainty. 
Other possibilities to determine these LECs considered in the
literature include fitting to the triton and $\alpha$-particle BEs
\cite{Nogga:2005hp} or the properties of light 
nuclei \cite{Navratil:2007we}. Recently, Gazit et al.~\cite{Gazit:2008ma} exploited
the fact that the LEC $c_D$ does not only contribute to the 3NF at N$^2$LO but also
governs the strength of the dominant short-range axial vector exchange
current and, therefore, can be determined from weak
processes. Using N$^3$LO 2NF of Ref.~\cite{Entem:2003ft} combined with N$^2$LO 3NF
with $c_{D,E}$ being calibrated to the triton BE and half-life time, the
authors of Ref.~\cite{Gazit:2008ma} obtain the $\alpha$-particle BE of
$28.50(2)$~MeV. Also, the point-proton radii of
$^3$H, $^3$He and $^4$He found to be $1.605(5)$~fm, $1.786(5)$~fm
and $1.461(2)$~fm  are in an excellent agreement with corresponding 
experimental values of $1.60$~fm, $1.77$~fm
and $1.467(13)$~fm, respectively. These results provide an important 
and highly nontrivial consistency
check of the chiral EFT approach by bridging the strong and
axial few-nucleon processes. 

Last but not least, the chiral 3NF at N$^2$LO is also being extensively
explored in connection with the spectra of light and medium-mass
nuclei, see \cite{Roth:2011ar} and Ref.~\cite{Navratil:2009ut} for a
recent review article describing state-of-the-art calculations within
the No Core Shell Model, the limit of neutron-rich nuclei
and, in particular, oxygen isotopes \cite{Otsuka:2009cs}, 
the properties of nuclear matter and constraints on
neutron star radii \cite{Hebeler:2010jx} and the puzzle of the 
anomalously long beta decay lifetime of $^{14}$C \cite{Holt:2010ma}.

\section{Nuclear lattice simulations}

A novel scheme to tackle the nuclear A-body problem that combines
chiral EFT for nuclear forces  
with Monte Carlo methods that are so successfully used in lattice QCD
or other fields of physics are the so-called ``nuclear lattice simulations'',
also named nuclear lattice EFT (NLEFT). In what follows, we will give a brief
outline of this approach and present some first results (for a
review with many references to earlier related work, see Ref.~\cite{Lee:2008fa}).

\subsection{Formalism}

\begin{figure}[t!]
\begin{center}
\includegraphics[width=6cm]{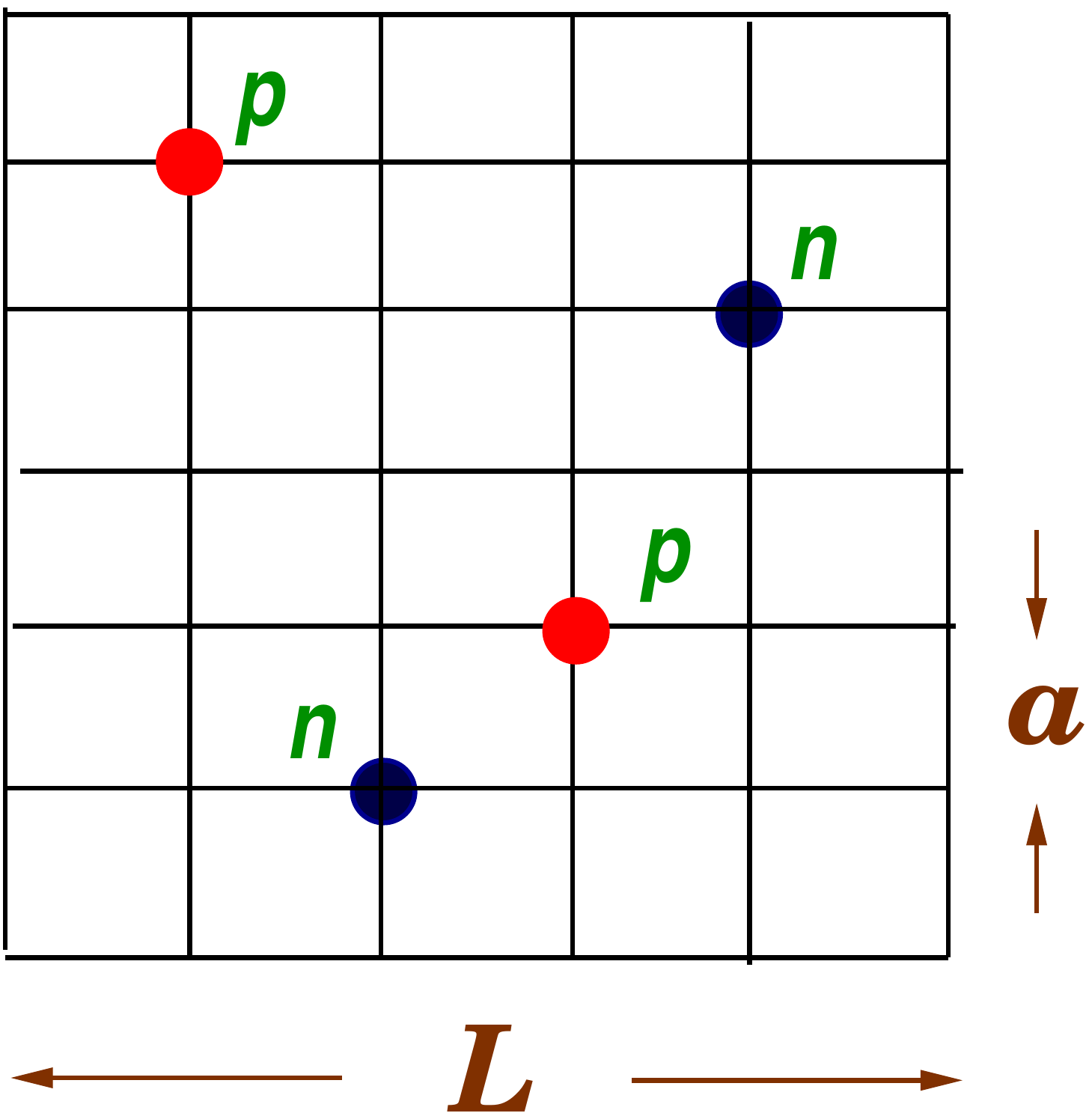}
\caption{\label{fig:lattice}
\baselineskip 12pt 
Schematic drawing of the space-time lattice. The minimal length is
the lattice spacing $a$. The side length $L$ in any spatial direction 
is an integer multiple of $a$. Protons ($p$) and neutrons ($n$) reside
on the lattice sites. 
}
\end{center}
\end{figure}
\begin{figure}[t!]
\begin{center}
\includegraphics[width=8cm]{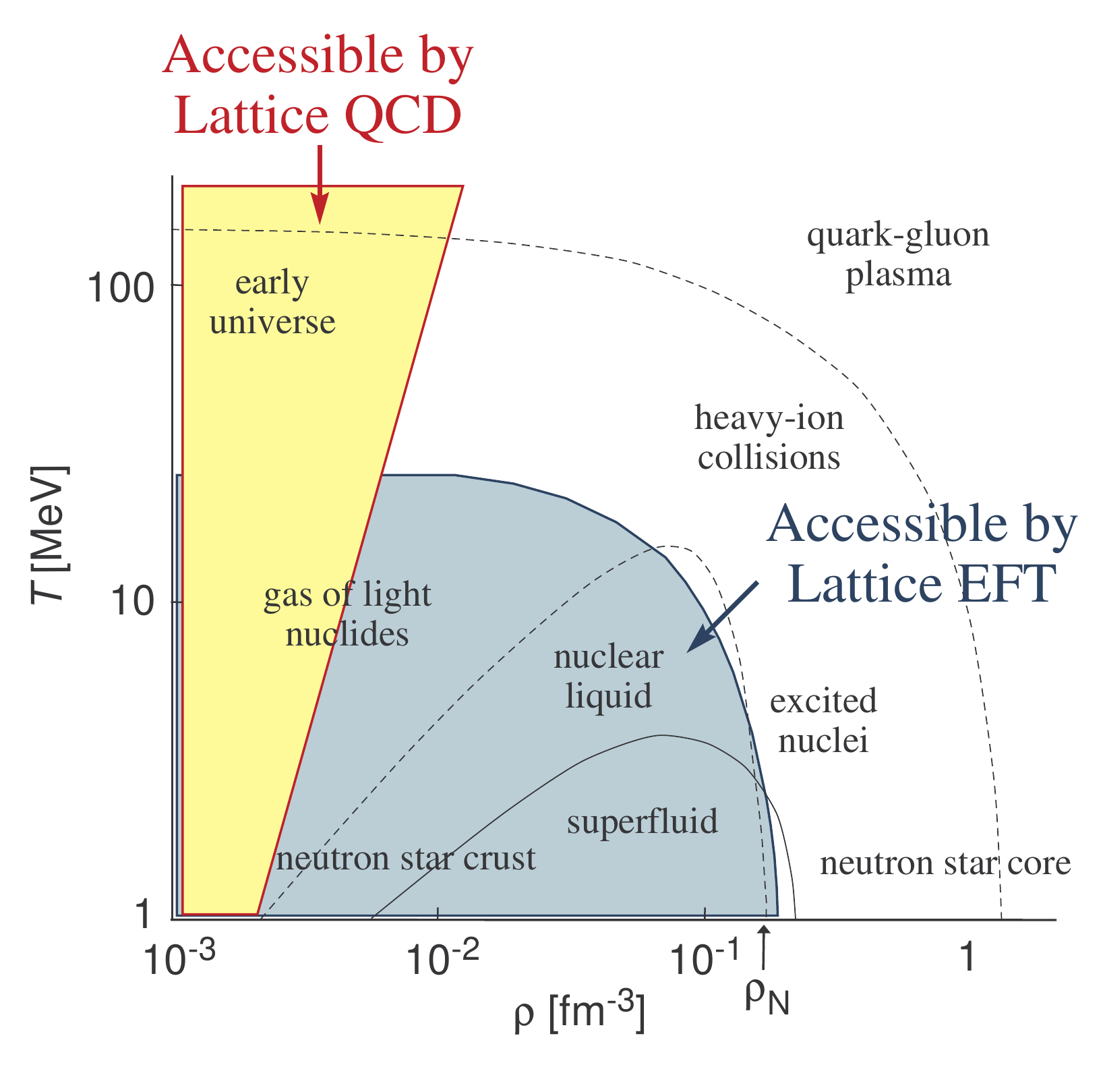}
\caption{\label{fig:phasedia}
\baselineskip 12pt 
Nuclear phase diagram as accessible by lattice QCD
(yellow area) and by nuclear lattice EFT (blue-grey area).
On the abscissa, the
nuclear density $\rho$ (with $\rho_N$ the density of
nuclear matter) and on the ordinate the temperature $T$
is displayed. Figure courtesy of Dean Lee.
}
\end{center}
\end{figure}

In NLEFT, space-time is discretized in Euclidean time on a torus of volume
$L_s\times L_s\times L_s\times L_t$, with $L_s (L_t)$ the side 
length in spatial (temporal) direction, as depicted in Fig.~\ref{fig:lattice}. 
The minimal distance on the lattice, the so-called lattice spacing, is $a$ ($a_t$)
in space (time). This entails a maximum momentum on the lattice,
$p_{\rm max} = \pi/a$, which serves as an UV regulator of the theory.
In contrast to lattice QCD, the continuum limit $a\to 0$ is not taken, as
we are dealing with an EFT and do not wish to resolve the structure of 
the individual nucleons. These are treated as point-like particles residing on the lattice sites,
whereas the nuclear interactions (pion exchanges and contact terms)
are represented as insertions on the nucleon world lines using standard
auxiliary field representations. The nuclear forces have an approximate
spin-isospin SU(4) symmetry (Wigner symmetry) \cite{Wigner:1936dx}
that is of fundamental importance in suppressing the malicious sign oscillations 
that plague any Monte Carlo (MC) simulation of strongly interacting fermion systems 
at finite density (for a modern look at this symmetry, see Ref.~\cite{Mehen:1999qs}).
The derivation of inequalities for binding energies of light nuclei in the Wigner symmetry limit
is given in Ref.~\cite{Chen:2004rq}. 
Because of this approximate symmetry, nuclear lattice simulations 
provide access to a large part
of the phase diagram of QCD, see Fig.~\ref{fig:phasedia}, whereas 
calculations using  lattice QCD are limited to finite temperatures and
small densities (baryon chemical potential). Here, we will concentrate
on the  calculation of the ground state properties and excited
states of atomic nuclei with atomic number $A \leq 12$.

We simulate the interactions of nucleons using the MC transfer matrix projection
method \cite{Montvay:1994cy}. Each nucleon evolves as if a single particle in a fluctuating
background of pion and auxiliary fields, the latter representing the
multi-nucleon contact interactions (for detailed definitions at LO
in the chiral expansion, see Ref.~\cite{Borasoy:2006qn}).
We also perform Gaussian smearing of the LO contact interactions 
which is required by the too strong binding of four nucleons on one lattice
site.  More precisely, in a LO calculation using the two independent 
four-nucleon contact operators without 
derivatives $\sim (N^\dagger N)^2$, one finds that in the $^4$He system the
ground state is severely overbound and consists almost entirely of the quantum
state with all four nucleons occupying the same lattice site. This is in part
due to a combinatorial enhancement of the contact interactions when more than two
nucleons occupy the same lattice site. This effect is partly overcome by higher
order four-nucleon operators, but it is most efficiently dealt with by
a Gaussian smearing procedure, which turns the point-like vertex into an
extended structure. A detailed discussion of this issue can be found 
in Ref.~\cite{Borasoy:2006qn}. Remarkably, the aforementioned configurations
lead to a new interpretation of the phenomenon of clustering in nuclei,
see Ref.~\cite{Meissner:2011cv}. 

\begin{figure}[t!]
  \centering
  \includegraphics[width=10.0cm]{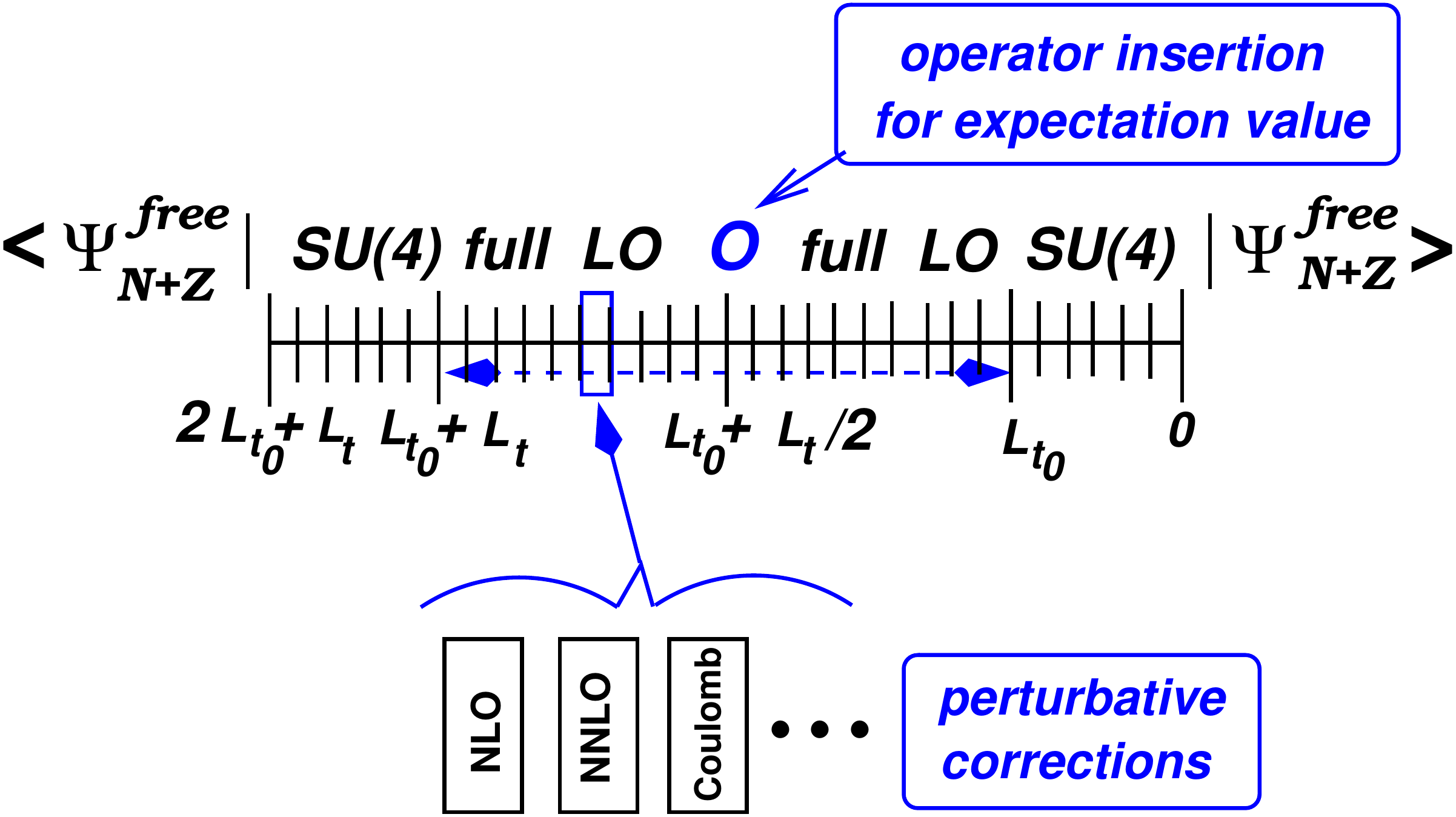}
  \caption{\baselineskip 12pt
  Schematic diagram for the transfer matrix calculation.  For the first
  ${t_0}$ time steps, the SU(4) symmetric part of the LO action is
  employed. This serves an inexpensive filter to suppress the sign oscillations.
  Only then the full LO action is used. All higher order corrections are
  included perturbatively as shown.  The initial and final wave function is
  a Slater determinant of $Z$ protons and $N$ neutrons.}
  \label{fig:time}
\end{figure}

Let us come back to the simulation method. To leading order, we start with a 
Slater determinant of single-nucleon standing waves in a periodic cube for 
$Z$ protons and $N$ neutrons (with $Z+N = A$). We use the SU(4) 
symmetric approximation of the LO interaction as an approximate 
inexpensive filter for the first $t_0$
time steps -- this suppresses dramatically the sign oscillations. Then we
switch on the full LO interaction and calculate the ground state energy and
other properties from the correlation function 
\begin{equation}
Z_A(t) = \langle \Psi_A |\exp(-tH)| \Psi_A \rangle~,
\label{eq:Z}
\end{equation} 
letting the Euclidean time $t$ go to infinity. 
Here $\Psi_A$ is the Slater-type initial wave function and $H$ is the
nuclear Hamiltonian, expressed in terms of the lattice variables and
lattice fields. Higher order
contributions, the Coulomb repulsion between protons and other
isospin-breaking effects (due to the light quark mass difference) are computed as
perturbative corrections to the LO transfer matrix. This is symbolically 
depicted in Fig.~\ref{fig:time}. The perturbative treatment of all these
effects is justified as for typical lattice
spacings $a \simeq 2\,$fm, the maximal momentum is $p_{\rm max} \simeq
300\,$MeV. 
Note, however, that due to the Gaussian smearing of the LO
contact interactions, part of the higher order corrections are also
treated non-perturbatively. If one is interested in the expectation value of
any operator ${\cal O}$, Eq.~(\ref{eq:Z}) has to be generalized as
\begin{equation}
Z_A^{\mathcal O} = \langle \Psi_A | \exp(-t H/2)\, {\cal O} \, \exp(-t H/2)\, |
\Psi_A\rangle~,
\end{equation} 
and the ground state expectation value is obtained as the Euclidean time tends
to infinity (see also Fig.~\ref{fig:time}).

Excited states are calculated from a multi-channel projection MC method,
briefly described in Ref.~\cite{Epelbaum:2011md}.
As a first step, we use various improvements in our LO lattice action.
This is necessary to suppress as much as possible the artefacts 
from the lattice, such that rotational symmetry $SO(3)$ is broken to cubic 
symmetry $SO(3,Z)$ and further artefacts due to the finite lattice
spacing $a$ arise. To minimize the effect of these, one performs
${\cal O}(a^4)$ improvements for the nucleon kinetic energy
and the Gaussian smearing factors of the contact interactions. Moreover,
all lattice operators at ${\cal O}(Q^3)$ are included, in particular
also the ones related to the breaking of rotational symmetry. Their
strengths can be tuned to eliminate unphysical partial wave mixing
like e.g. between the $^3S_1 - ^3D_1$ and the $^3D_3$ partial waves.
One then uses a set of Slater determinants of a large number of different single 
standing nucleon waves. E.g. for the calculation of the spectrum of
$^{12}$C, 24 initial standing waves were used and
from these  three states with total momentum zero and $J_z =0$ mod
4 and one state with  $J_z =2$ mod 4 were constructed. 
Therefore, the correlation function turns into a matrix, 
\begin{equation}
Z^{ij}_A (t) = \langle \Psi_A^i |\exp(-tH)| \Psi_A^j\rangle~.
\end{equation} 
Diagonalization of this matrix of a given ensemble of states with
the required quantum numbers 
leads to a tower of states. In such a way, one
is able to reconstruct the excitation spectrum of any given nucleus. However,
due to the required computing resources, the ground state and only a few 
excited states for some nuclei have been computed so far.

The very low memory and trivially parallel structure of the lattice MC codes 
allows to perform simulations, that scale ideally with several thousands
of processors. The computational time scales with the number of nucleons $A$ as $A^{1.7}$
at fixed volume $V$ and with $V^{1.5}$  for fixed $A$. The average sign --
which is  a measure of the severity of the sign oscillations -- scales
approximately as $\exp(-0.1A)$. Taking the calculation of $^{12}$C as a
benchmark, the required CPU time for a nucleus with spin $S$ and isospin
$I$ can be estimated as
\begin{equation}
\label{eq:cpugen} 
X^{\rm CPU} \approx X_{^{12}C}^{\rm CPU} \times \left(
  \frac{A}{12}\right)^{3.2}\exp[0.1(A-12) + 3(S~{\rm mod}~2) +4I]~,
\end{equation} 
and the memory requirements to store the generated configurations are
\begin{equation} 
\label{eq:stogen} 
X^{\rm storage} \approx X_{^{12}C}^{\rm storage} \times \left(
  \frac{A}{12}\right)^{2}\exp[0.1(A-12) + 3(S~{\rm mod}~2) +4I]~.
\end{equation} 
Therefore, combining high performance computing with the forces
derived from chiral effective field theory and fixing the parameters
in  few-nucleon systems allows for true {\sl ab initio} calculations
of atomic nuclei and their structure, with a quantifiable uncertainty
of any observable under investigation. Before presenting first result
based on nuclear lattice EFT, it is important to stress the differences to other
{\sl ab initio} methods. One distinction is that in NLEFT all systematic
errors are introduced up front when defining the low-energy EFT. This
eliminates unknown approximation errors related to specific calculational
tools, physical systems or observables. By including higher-order interactions
one can expect systematic improvement for all low-energy observables.
Another difference is that many different phenomena can be studied using the
same lattice action. Once the action is determined, it can be used to
calculate bound nuclei, the ground state of neutron matter or thermodynamic
properties at non-zero temperature. In addition, NLEFT is utilizing several
efficient lattice methods developed for lattice QCD and condensed matter
simulations, including Markov Chain MC techniques, 
auxiliary fields \cite{Hubbard:1959ub,strato}, pseudofermion methods 
\cite{Weingarten:1980hx} and non-local updating schemes such as hybrid MC
\cite{Scalettar:1986uy,Gottlieb:1986ms,Duane:1987de}.

\subsection{Results}

\begin{figure}[t!]
  \centering
  \includegraphics[width=10.0cm]{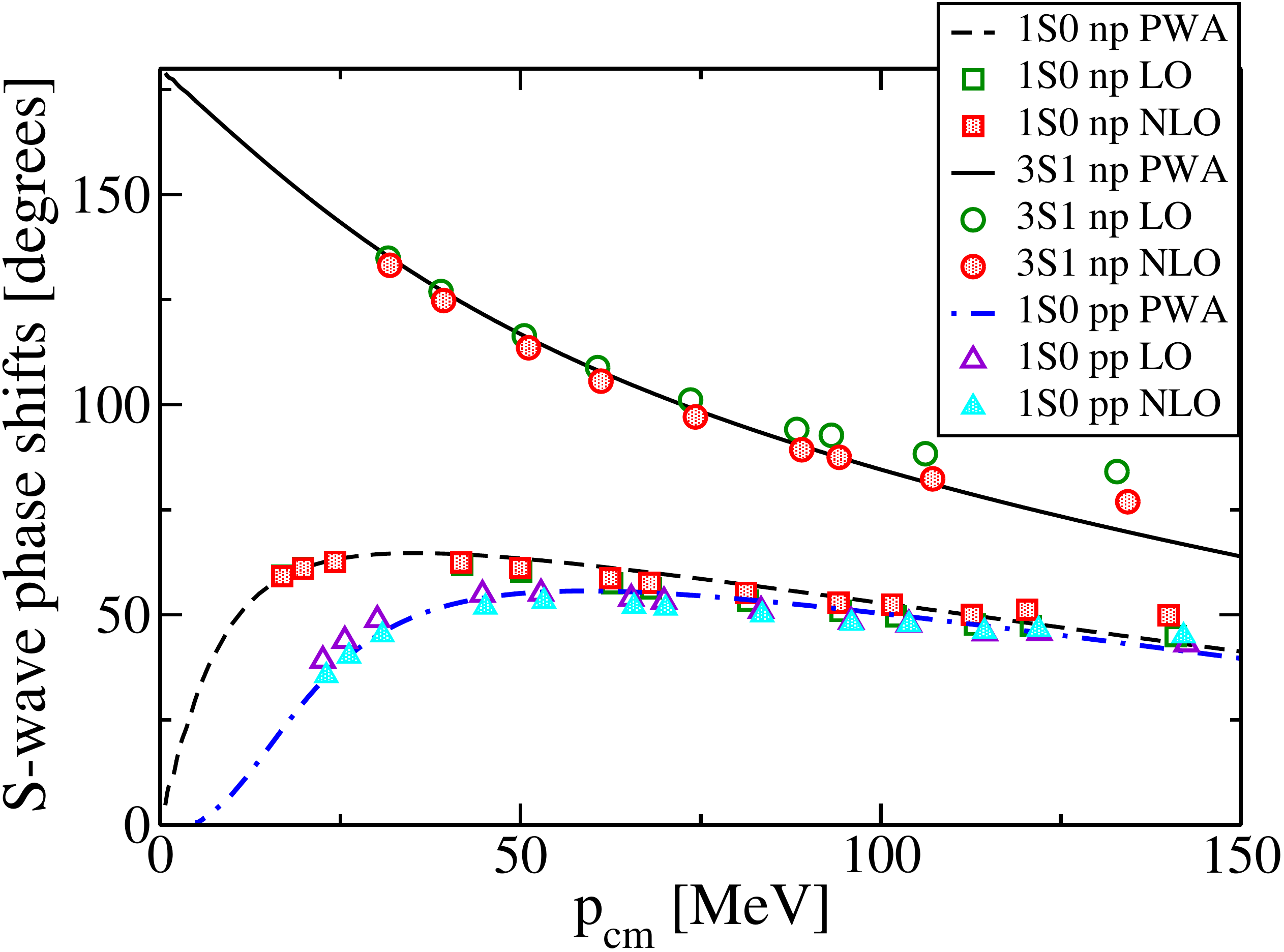}
  \caption{
    \baselineskip 12pt 
  S-wave phase shifts in the two-nucleon system. From top to bottom: $^3S_1$
  (np), $^1S_0$(np), $^1S_0$(pp). The open/filled symbols represent the results
  at LO/NLO (including all isospin-breaking effects), respectively. The
  curves are the result of the Nijmegen PWA. While the $np$ phases
  are fitted, for the $pp$ phase only the proton-proton scattering length is
  input.}
  \label{fig:Swaves}
\end{figure}

So far, calculations in NLEFT have been performed up to next-to-next-to-leading order 
in the chiral expansion of the nuclear potential. At this order, the
two- as well as the leading three-body forces are present. Consider first the two-nucleon
system. At NLO, we have nine parameters that are determined from a fit to the 
S- and P-waves in $np$ scattering. Two further
isospin-breaking parameters are determined from the $pp$ and $nn$ scattering
lengths \cite{Borasoy:2007vi}. As shown in Fig.~\ref{fig:Swaves}, up to
center-of-mass momenta of the order of the pion mass, the empirical
phase shifts in the neutron-proton system are well described and, 
furthermore, the NLO corrections are small. 
To arrive at these results, a novel method
to extract phase shifts from finite volume simulations had to be developed
(as the standard L\"uscher scheme is not well suited in case of strong partial
wave mixing) \cite{Borasoy:2007vy}. Note that due to the perturbative
treatment of the higher-order effects, there are no contributions to the 2NF
at N$^2$LO.
The three-nucleon force features only two LECs, see the discussion in
section \ref{sec_forces},
that are determined in our simulations from the triton binding energy
in combination either with low-energy 
neutron-deuteron scattering in the doublet channel
\cite{Epelbaum:2009zsa} or the $\alpha$-particle BE \cite{Epelbaum:2011md} .

The first non-trivial predictions are i) the energy dependence of the $pp$
$^1S_0$ partial wave, which agrees with the Nijmegen partial wave analysis up to momenta
of about the pion mass, see the lowest curve and triangle symbols 
in Fig.~\ref{fig:Swaves}, and ii) 
the binding energy difference between the
triton ($^3$H) and $^3$He,
\begin{equation}
E(^3{\rm He}) - E(^3{\rm H}) = 0.78(5)~{\rm MeV}~, 
\end{equation}
in good agreement with the experimental value of
$0.76$~MeV \cite{Epelbaum:2009pd,Epelbaum:2010xt}. The theoretical
uncertainty mostly arises from the infinite-volume extrapolation.
To arrive at the result for the tri-nucleon ground state energies, 
the finite volume expression
\begin{equation}
E(L) = -{\rm BE} -
\displaystyle\frac{a}{L} \exp(-bL)~,
\end{equation}
where BE denotes the (positive) binding energy, 
$E(L)$ the measured energy
in the finite volume $L^3$ and $a,b$ are fit parameters, has been
utilized (see also Ref.~\cite{Kreuzer:2010ti}).

\begin{figure}[t!]
  \centering
  \includegraphics[width=8.0cm]{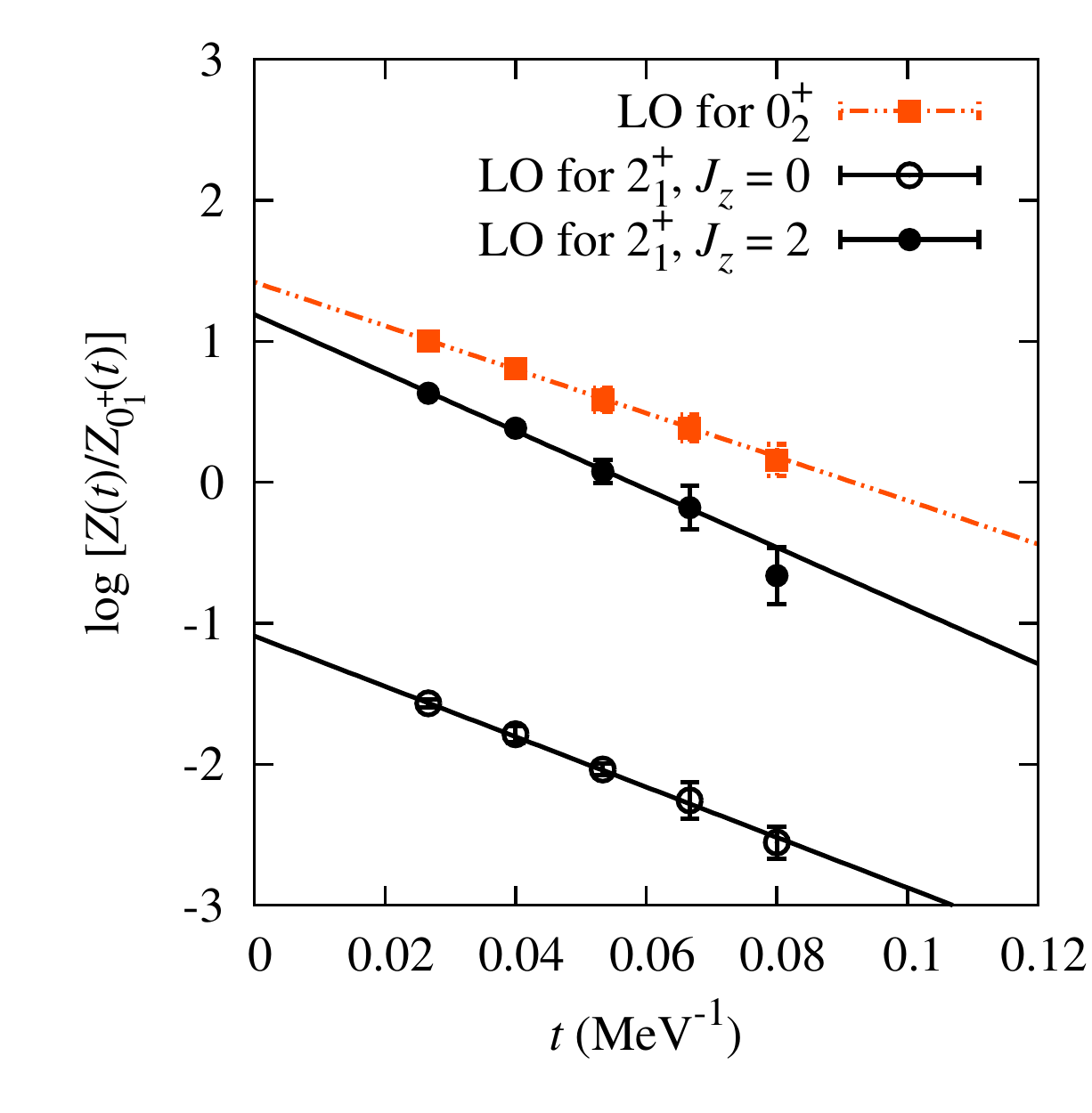}
  \caption{
    \baselineskip 12pt 
     Extraction of the excited states of $^{12}$C from the Euclidean
      time dependence
      of the projection amplitude at LO. \ The slope of the logarithm of
      $Z(t)/Z_{0_{1}^{+}}(t)$ at large $t$ determines the energy relative to the
       ground state.}
  \label{fig:12Cexcited}
\end{figure}

The ground state energies of nuclei with $A = 4,6,12$ where  calculated in 
Refs.~ \cite{Epelbaum:2009pd,Epelbaum:2010xt}, showing that at N$^2$LO one
can achieve a precision of a few percent. Refining the underlying action
as described above and utilizing the multi-channel projection MC method,
the spectrum of $^{12}$C was worked out in \cite{Epelbaum:2011md}.
In Fig.~\ref{fig:12Cexcited}, we show the clean signals of the first 
few excited states
on top of the $0^+$ ground state. Note that due to lattice artefacts the
first excited $2^+$ state is indeed split into two states. This will
eventually be overcome by choosing a larger basis of initial states.
In addition to the ground state and the
excited spin-2 state, the calculation
gives a resonance with angular momentum zero and positive parity at
$-85(3)$~MeV, very close to the $^4$He+$^8$Be threshold at $-86(2)$~MeV.
Experimentally this threshold is located at $-84.80$~MeV.
This first $0^+$ excitation is the so-called Hoyle state. 
It plays a crucial role in the helium burning 
of stars heavier than our sun and in the production of carbon and
other elements necessary for life. This excited state of $^{12}$C was
postulated  by Hoyle \cite{Hoyle:1954zz} as a necessary ingredient for 
the fusion of three $\alpha$-particles to produce a sufficient amount of 
carbon and other elements needed for life at stellar
temperatures. For this reason, the Hoyle state plays also a very important 
role in the context of the anthropic principle, although such considerations 
did not play any role when this state was predicted \cite{Kragh}. The Hoyle
state has been an enigma for nuclear structure theory since decades, even
the most successful Greens function MC methods based on realistic two- and
three-nucleon forces \cite{Pieper:2007ax} or the no-core-shell-model 
employing modern (chiral or $V_{\rm low~ k}$) interactions
\cite{Navratil:2007we,Roth:2011ar} have not been able to describe this state.
\renewcommand{\arraystretch}{0.8}
\begin{table}[t!]
\begin{center}
\begin{tabular}
[c]{|l||c|c|c|c|}\hline
& $0_{1}^{+}$ &$0_{2}^{+}$ & $2_{1}^{+}$, $J_{z}=0$ & $2_{1}^{+}$, $J_{z}=2$\\\hline\hline
LO [$O(Q^{0})$] & $-110(2)$ & $-94(2)$ & $-92(2)$ & $-89(2)$\\\hline
NLO [$O(Q^{2})$] & $-85(3)$ & $-74(3)$ & $-80(3)$ & $-78(3)$\\\hline
N$^2$LO [$O(Q^{3})$] & $-91(3)$ & $-85(3)$ & $-88(3)$ & $-90(4)$\\\hline
Experiment & $-92.16$ & $-84.51$ & \multicolumn{2}{|c|}{$-87.72$}\\\hline
\end{tabular}
\end{center}
\caption{\baselineskip 12pt
Lattice results for the ground state $0_{1}^{+}$
and the low-lying excited  states of $^{12}$C. For comparison 
the experimentally observed energies are shown. All energies are in units of MeV.}
\label{excited states}%
\end{table}
In Table~\ref{excited states} we show results for the ground state and
the low-lying excited states of $^{12}$C at LO,
NLO  with isospin-breaking and electromagnetic corrections
included, and N$^2$LO.  
For comparison we list the experimentally observed energies.
The error bars in Table~\ref{excited states} are one standard
deviation estimates which include both Monte Carlo statistical errors and
uncertainties due to extrapolation at large Euclidean time. \ Systematic
errors\ due to omitted higher-order interactions can be estimated from the
size of corrections from $O(Q^{0})$ to $O(Q^{2})$ and from $O(Q^{2})$ to
$O(Q^{3})$.  
\begin{figure}[t!]
  \centering
  \includegraphics[width=8.0cm]{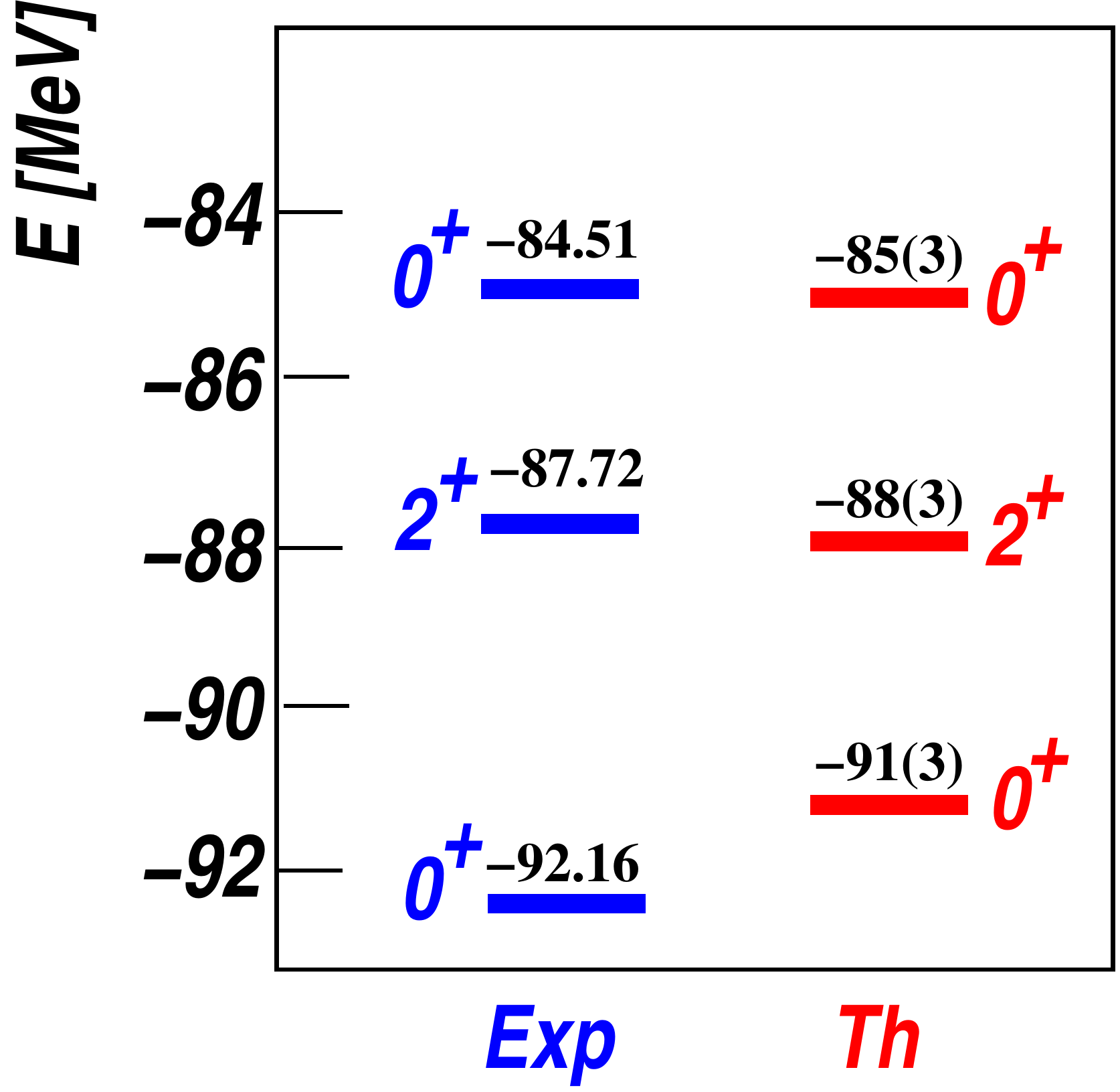}
  \caption{
  \baselineskip 12pt
  Results for the  $^{12}$C spectrum (Th) and comparison with  
  experiment (Exp). The results of the NLEFT at N$^2$LO are shown for
  the ground state, the Hoyle state and the 
  lowest-lying spin-2 state. In the simulations, the $2^+$ state
  is split into the $J_z =0$ and  
  $J_z =2$  projections. Only the $J_z =0$ component is shown here.}
  \label{fig:12Cspectrum}
\end{figure}
As seen in Table~\ref{excited states} and summarized in 
Fig.~\ref{fig:12Cspectrum}, the N$^2$LO results for the Hoyle state
and spin-2 state are in agreement with the experimental values.  While the
ground state and spin-2 state have been calculated in other studies, these 
results are the first \textit{ab initio} calculations
of the Hoyle state with an energy close to the phenomenologically important
$^{8}$Be-alpha threshold. It is important to  note the energy level 
crossing involving the Hoyle state and the spin-2
state.  The Hoyle state is lower in energy at LO but higher at NLO. One of
the main characteristics of the NLO interactions is to increase the repulsion
between nucleons at short distances.  This has the effect of decreasing the
binding strength of the spinless states relative to higher-spin states.  We
note the $25$~MeV reduction in the ground state binding energy and $20$~MeV
reduction for the Hoyle state while less than half as much binding
correction for the spin-2 state.  This degree of freedom in the energy
spectrum suggests that at least some fine-tuning of parameters is needed to
set the Hoyle state energy near the $^{8}$Be+$^4$He threshold.  It would be
very interesting to understand which fundamental parameters in nature control
this fine-tuning.  At the most fundamental level there are only a few such
parameters, one of the most interesting being the masses of the up and down
quarks. Such investigations have already been performed to unravel the quark
mass dependence of the deuteron binding energy and of the S-wave
nucleon-nucleon scattering lengths \cite{Beane:2002vs,Epelbaum:2002gb}.
The impact on the primordial abundances of light elements created by a 
variation of the quark masses at the time of Big Bang nucleosynthesis was
also studied in Ref.~\cite{Bedaque:2010hr}.

\subsection{Neutron matter}

Matter purely made of neutrons is not only interesting by itself but also
of astrophysical relevance as various  forms of this strongly interacting
quantum many-body state are realized in the different layers of neutron stars.
Due to the Pauli-principle, three-body forces are suppressed in neutron matter
and, of course, there is also no Coulomb repulsion. As for nuclei, one can
perform simulations for a fixed number of neutrons in a given volume, thus
varying the Fermi momentum $k_F$ (density $\rho$) of the neutron matter. 
For $N$ (spin-saturated)  neutrons in a periodic cube of side-length $L$, 
the Fermi momentum (density) is
\begin{equation} 
\label{eq:dens} 
k_F = \displaystyle\frac{(3\pi^2N)^{1/3}}{L}~,~~ 
\rho = \displaystyle\frac{k_F^3}{3\pi^2}~. 
\end{equation} 
Varying $L$ from 4 to 7 and  $N = 8,12,16$  corresponds to a Fermi momentum
between $88$ and $155$~MeV, i.e. densities between 2 and 10\% of normal 
nuclear matter density, $\rho_N = 0.17\,$fm$^3$. Interestingly, 
neutron matter at $k_{F}\sim 80$~MeV is close to the so-called {\sl unitary limit},
where the $S$-wave scattering length is infinite and the range of the
interaction is negligible.  At lower densities corrections due to the
(finite) scattering length become more important, and at higher densities corrections
due to the effective range and other effects become important.  In the
unitary limit the ground state has no dimensionful parameters other than
the particle density. Thus, the ground state energy of the system should obey the
simple relation $E_{0}=\xi E_{0}^{\text{free}}$ for some dimensionless
constant $\xi$, with $E_{0}^{\text{free}}$ the energy of non-interacting particles.
The universal nature of the unitary limit endows it
relevance to several areas of physics, and in atomic physics the unitarity
limit has been studied extensively with ultracold $^{6}$Li and $^{40}$K atoms
using a magnetic-field Feshbach resonance (a summary of recent determinations
of $\xi$ can be found in Ref.~\cite{Bour:2011xt}). 
There have been numerous analytic calculations of
$\xi$, employing the full arsenal of available many-body techniques, see
Ref.~\cite{Furnstahl:2008df} for a recent review. A benchmark calculation 
for the four-particle system was reported in Ref.~\cite{Bour:2011xt}.
Around the unitary limit,
the ratio  $E_{0}/E_{0}^{\text{free}}$ can be parameterized as
\begin{equation} 
\label{eq:unilim} 
\displaystyle\frac{E_{0}}{E_{0}^{\text{free}}} =  \xi - \frac{\xi_1}{k_F\,
  a_{nn}} + \xi_2 \, k_F \, r_{nn} + \ldots~
\end{equation} 
in terms of the neutron-neutron scattering length $a_{nn}$ 
and effective range $r_{nn}$. In what follows, we use $\xi = 
0.31(1)$ and $\xi_1 = 0.81(1)$ as determined
from lattice simulations for two-component fermionic systems
in Ref.~\cite{Lee:2007ae,Lee:2008xsa}.

\begin{figure}[t!]
  \centering
  \includegraphics[width=10.0cm]{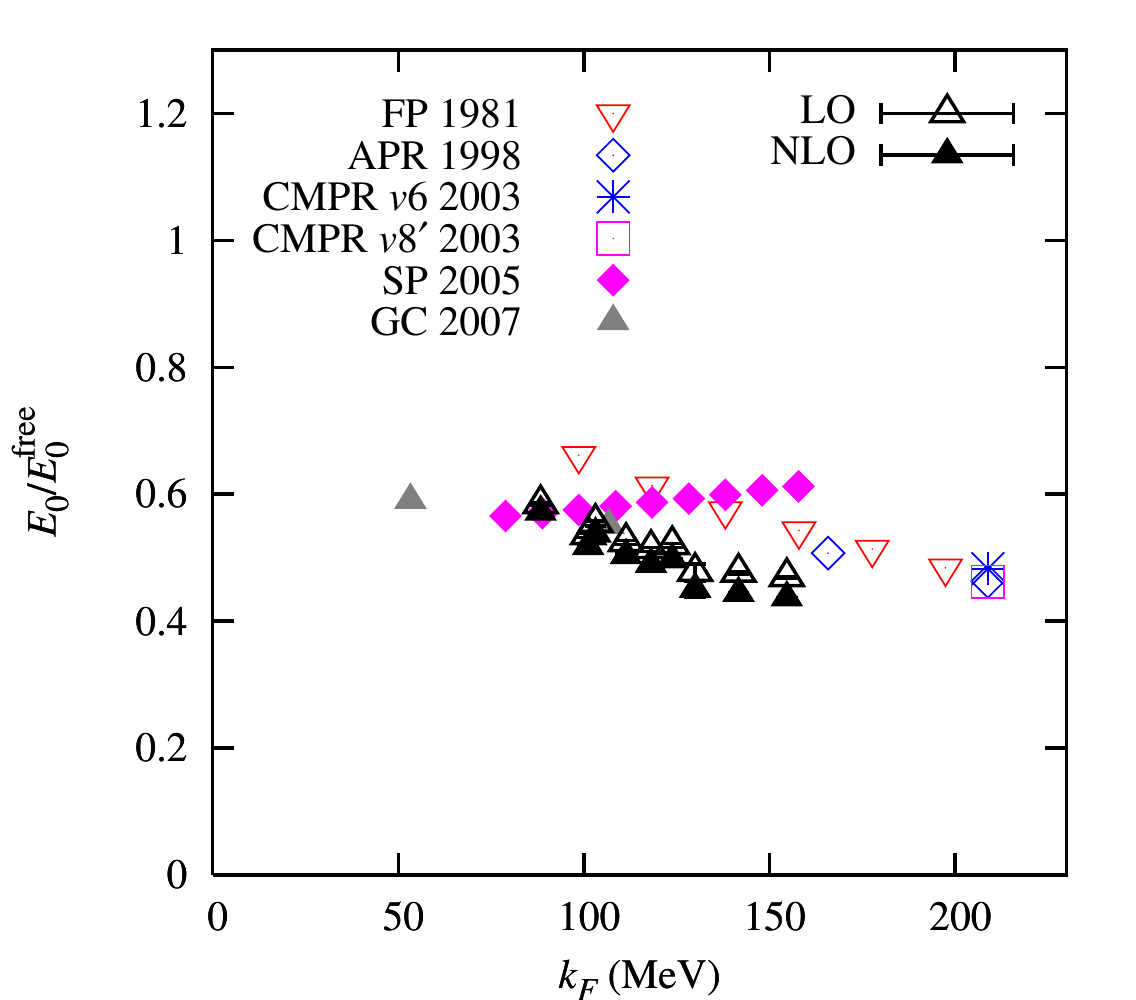}
  \caption{
  \baselineskip 12pt
  Results for ground state of strongly interacting neutron matter 
  $E_0/E_0^{\rm free}$ versus the Fermi momentum $k_F$ using NLEFT 
  at LO (open triangles) and NLO (filled triangles). For comparison, 
  the results of
  FP~1981~\cite{Friedman:1981qw}, APR~1998~\cite{Akmal:1998cf},
  CMPR~$v6$ and $v8^\prime$~2003~\cite{Carlson:2003wm},
  SP~2005~\cite{Schwenk:2005ka} and GC~2007~\cite{Gezerlis:2007fs}
  based on different many-body techniques are also displayed.}
  \label{fig:nmatter}
\end{figure}

In the framework of NLEFT, neutron matter was studied in 
Refs.~\cite{Borasoy:2007vk,Epelbaum:2008vj}. The resulting
energy of the $N$ neutron quantum state compared to a free
ensemble is shown in Fig.~\ref{fig:nmatter}, in comparison
to earlier calculations using different many-body techniques. 
For most cases in the range of densities considered, the
agreement is good. The universal parameter $\xi_2$ from 
Eq.~(\ref{eq:unilim}) is determined to be in the range from 
0.14 to 0.27~\cite{Epelbaum:2008vj}. In principle,
it can be measured in any two-component fermionic system.

In the future, nuclear lattice simulations of  neutron matter can
be used to investigate the interesting problem of a possible
P-wave pairing, as we can dial the strength of the neutron-neutron
interactions in the various partial waves by varying the strength
of the corresponding LECs. Obviously, this
is not possible in nature. Also, simulations for more neutrons
and different lattice spacings are necessary to get a better grip
on the neutron equations of state (EoS) at higher densities and also
get a better control of the lattice errors. With the observation
of a neutron star of two solar masses \cite{Demorest:2010bx}, there exist now much
more stringent constraints on the neutron EoS (see e.g. 
Refs.~\cite{Steiner:2010fz,Hebeler:2010jx})
and therefore such investigations based on NLEFT will be an 
important tool.

\section*{Acknowledgements} 
We thank all our collaborators for sharing their insights into topics
discussed here. We are also very grateful to V\'eronique Bernard,
Hermann Krebs and Dean Lee  for useful comments on the manuscript. 
This work is partly supported by the Helmholtz Association through the
Nuclear Astrophysics Virtual Institute (VH-VI-417), 
by the EU HadronPhysics3 project ``Study of strongly interacting matter'', 
by the European Research Council (ERC-2010-StG 259218 NuclearEFT) and 
by the DFG (TR 16, ``Subnuclear Structure of Matter''). 

\end{document}